\documentclass[11pt,letterpaper,english]{article}

\usepackage{mathptmx}
\usepackage{microtype}
\usepackage{float}
\usepackage{amsmath} 
\usepackage{amsthm}
\usepackage{mathrsfs}
\usepackage{bm}
\usepackage{bbm}
\usepackage{graphicx} 
\usepackage{caption}
\usepackage{setspace}
\usepackage{amssymb}
\usepackage{multirow}
\usepackage{rotating}
\usepackage{array}
\usepackage{booktabs}
\usepackage{multibib}
\usepackage{bibunits}
\usepackage{xr}
\usepackage{algorithm}
\usepackage{algpseudocode}

\usepackage[normalem]{ulem}

\usepackage{algpseudocode}
\usepackage{color}
\usepackage{pifont}
\usepackage{threeparttable}
\usepackage{booktabs}
\usepackage{tabularx}  
\usepackage[margin=1in]{geometry}  

\usepackage[small,compact]{titlesec} 
\DeclareMathAlphabet{\mathcalligra}{T1}{calligra}{m}{n}
\usepackage[sectionbib,round]{natbib}
\usepackage{times}
\usepackage{abstract}

\usepackage[hang,flushmargin]{footmisc}
\usepackage{fullpage} 
\usepackage{hyperref}
\usepackage[compact]{titlesec}
\usepackage[section]{placeins}
\usepackage{footnote}
\usepackage{placeins}
\usepackage[toc,page]{appendix}
\usepackage{subcaption}
\usepackage{blkarray}
\usepackage{stackengine}
\usepackage{accents}
\usepackage{enumitem}
\usepackage[mathscr]{euscript}

\defaultbibliographystyle{abbrvnat} 
\defaultbibliography{ref} 

\usepackage{xcolor}
\usepackage[small,compact]{titlesec} 
\usepackage[toc,page]{appendix}
\usepackage{bibunits}
\defaultbibliographystyle{abbrvnat} 
\defaultbibliography{ref} 
\usepackage{listings}
\bibpunct[, ]{(}{)}{;}{a}{}{,}

\theoremstyle{definition}

\theoremstyle{definition}

\usepackage[utf8]{inputenc}
\usepackage{longtable}
\usepackage{epstopdf}
\usepackage{bbm}
\usepackage{dsfont}
\usepackage{comment}
\usepackage{pgfplots}
\pgfplotsset{compat=1.18}
\definecolor{englishblue}{HTML}{1F77B4}
\definecolor{germanorange}{HTML}{FF7F0E}

\usepackage{pifont}%

\usepackage{caption}
\newcommand{\paperassetgraphic}[2][]{%
  \IfFileExists{#2}{%
    \includegraphics[#1]{#2}%
  }{%
    \fbox{%
      \begin{minipage}[c][0.055\textheight][c]{0.92\linewidth}
      \centering\scriptsize Missing figure
      \end{minipage}%
    }%
  }%
}
\newcommand{\paperemptyassetgraphic}[2][]{%
  \IfFileExists{#2}{%
    \includegraphics[#1]{#2}%
  }{%
    \makebox[\linewidth][c]{%
      \fbox{%
        \begin{minipage}[c][0.058\textheight][c]{0.45\linewidth}
        \centering\scriptsize non-PSD cov. matrix
        \end{minipage}%
      }%
    }%
  }%
}

\theoremstyle{plain}

\newcommand{\squishlist}{
   \begin{list}{$\bullet$}
    { \setlength{\itemsep}{0pt} \setlength{\parsep}{1pt}
      \setlength{\topsep}{1pt} \setlength{\partopsep}{1pt}
      \setlength{\leftmargin}{1.5em} \setlength{\labelwidth}{1em}
      \setlength{\labelsep}{0.5em} } }

\newcommand{\squishlisttwo}{
   \begin{list}{$\bullet$}
    { \setlength{\itemsep}{0pt} \setlength{\parsep}{0pt}
      \setlength{\topsep}{0pt} \setlength{\partopsep}{0pt}
      \setlength{\leftmargin}{1em} \setlength{\labelwidth}{1.5em}
      \setlength{\labelsep}{0.5em} } }

\newcommand{\squishend}{
    \end{list}  }

\begin{document}
\newcommand{\clickMainDEPairs}{499{,}927}
\newcommand{\clickMainDEBeta}{-0.0561}
\newcommand{\clickMainDEPValue}{0.0004}
\newcommand{\clickMainDEDecline}{5.45\%}
\newcommand{\clickMainFRPairs}{530{,}873}
\newcommand{\clickMainFRBeta}{-0.0494}
\newcommand{\clickMainFRPValue}{0.0035}
\newcommand{\clickMainFRDecline}{4.82\%}
\newcommand{\clickMainJAPairs}{315,054}
\newcommand{\clickMainJABeta}{-0.1806}
\newcommand{\clickMainJAPValue}{0.0000}
\newcommand{\clickMainJADecline}{16.53\%}
\newcommand{\clickMainEconomicBeta}{-0.0527}
\newcommand{\clickMainEconomicDecline}{5.14\%}
\newcommand{\clickMainFRPostAverageBillions}{1.852}
\newcommand{\clickMainMonthlyLossMillions}{100.27}
\newcommand{\clickMainAnnualLossBillions}{1.20}
\newcommand{\clickMainMonthlyRevenueLowMillions}{0.901}
\newcommand{\clickMainMonthlyRevenueHighMillions}{3.090}
\newcommand{\clickMainAnnualRevenueLowMillions}{10.82}
\newcommand{\clickMainAnnualRevenueHighMillions}{37.08}

\title{Impact of AI Search Summaries on Website Traffic: Evidence from Google AI Overviews and Wikipedia}

\author{
Mehrzad Khosravi \thanks{ Please address all correspondence to: mehrzad@uw.edu and hemay@uw.edu.}\\ University of Washington
\and
Hema Yoganarasimhan\\
University of Washington
}


\maketitle

\begin{abstract}
\begin{singlespace}
Search engines increasingly display AI-generated answers above organic links, potentially displacing traffic to upstream publishers. We estimate the impact of Google’s AI Overviews (AIO) on Wikipedia's search traffic using AIO's staggered geographic rollout and Wikipedia’s multilingual structure. Our difference-in-differences design compares monthly external-search referrals to English Wikipedia articles with referrals to the same articles in German and French, and finds that default AIO availability reduced English search traffic by \clickMainDEDecline\ and \clickMainFRDecline\,, respectively. Our results suggest that answer-producing digital intermediaries can materially reallocate attention away from informational publishers, with implications for content monetization, search platform design, and policy.

\end{singlespace}
\end{abstract}

\noindent \textbf{Keywords:} Generative AI, Search Engines, Digital Economy, Artificial Intelligence, Wikipedia, Google

\newpage

\begin{bibunit}

\section{Introduction}
\label{sec:introduction}

Digital information intermediaries, such as search engines and content
aggregators, have traditionally created value by organizing third-party content, lowering search costs, and directing users to relevant publishers. Generative AI changes this role by allowing search engines and information intermediaries to synthesize information from multiple upstream publishers and present a self-contained response within their own interfaces. Rather than merely helping users locate relevant content, these intermediaries can now satisfy part of users' information demand before they visit the underlying source. This shift can improve access to information and further reduce search costs, but it may also divert attention and economic value from the publishers whose content supports the generated response. Because many publishers rely heavily on search referrals, even modest reductions in referral traffic can affect advertising revenue, subscription acquisition, and ultimately the incentives to produce and maintain content \citep{Nielsen2022Concentration}. Whether answer-producing information intermediaries complement or substitute for upstream content producers is therefore an important question for the digital economy. 

Google's AI Overviews (AIO) provides a prominent setting to study this question. Google's AIO feature presents a synthesized answer prominently on the search-results page, before the traditional organic links, and includes links to a limited
set of cited sources
\citep{Google2024SuperChargeSearch,Google2024GenerativeAISearch}; see Web Appendix $\S$\ref{appsec:serps} for an example. Google made
AIO part of the default search experience for U.S.\ users in May 2024, \citep{Google2024GenerativeAISearch}, and then expanded AI Overviews internationally in subsequent months, reaching more than 100 countries by early  2025 \citep{Google2024AIOExpansion}. The change is economically consequential because Google accounts for approximately $90\%$ of the global search market \citep{StatistaSearchShare}. 

There is substantial disagreement over the effect of AIO on upstream publisher traffic. Publishers argue that by placing a synthesized answer above the organic results, AIO may satisfy users' information needs without an external click and push source pages farther down the search-results page. In this view, AIO acts as a \emph{substitute} for publisher content and cannibalizes traffic, even when it cites the underlying sources \citep{Digiday2025AIO, Wikimedia2025traffic}. Google counters that AIO is \emph{complementary}: it lowers search costs, improves user satisfaction, and sends users to helpful pages through included links, potentially expanding overall search activity and traffic to publishers \citep{Google2024GenerativeAISearch, Google2025AISearchClicks}. 


The economic and policy stakes of this disagreement are substantial. U.S.\
search advertising generated \$102.9 billion in revenue in 2024, accounting
for nearly 40\% of all U.S.\ internet advertising revenue
\citep{IAB2025AdRevenue2024}. Google has also steadily integrated advertising
into AIO, expanding these placements across devices and geographies by late
2025 \citep{Google2025AdsInAIO}. These developments increase the economic
importance of where users consume information: when AIO satisfies a query
within the search interface, the platform can retain and monetize the user's
attention while the underlying publishers may lose visits and associated
revenue opportunities. If AIO systematically diverts traffic from source
pages, it could shift advertising revenue, user-interaction data, bargaining
power, and ultimately market power from upstream publishers toward large
search platforms. Over time, lower traffic and monetization could weaken
publishers' incentives to create and maintain the content on which generative
search relies, potentially generating a negative feedback loop in content
supply. If, instead, AIO lowers search costs and directs users toward more
relevant or higher-quality sources, it may improve the efficiency of
information search and generate more valuable visits for publishers.
Distinguishing between these possibilities is therefore central to
understanding how generative search redistributes attention and economic value, and how it shapes the long-run incentives to produce the content on which it depends.


At a high-level, this debate echoes prior work on news aggregators, which asks whether aggregators cannibalize publisher demand (substitution) or expand it by improving discovery and lowering search costs (complementarity). Empirical evidence largely supports complementarity: removing Associated Press content from Google News reduced traffic to other publishers \citep{chiou2017content}, and the unavailability of Google News in Spain decreased traffic to Spanish publishers \citep{calzada2020news, athey2021impact}. However, AIO changes the mechanism in a way that makes direct extrapolation difficult. Unlike an aggregator that primarily routes users to linked pages, AIO presents a synthesized answer before organic links and can satisfy informational queries without a click. Whether this shift from routing demand to fulfilling it turns the intermediary from a complement into a substitute is therefore an empirical question.


We estimate the causal effect of Google’s AI Overviews on publisher search traffic, using Wikipedia as a large, transparent, and economically important empirical setting. Wikipedia is particularly well suited to this analysis for three reasons. First, it is one of the world's most visited websites and a major source of informational content, receiving more than 130 billion page views per year \citep{TopVisited2024, WikipediaStats2024}. Search engines also play a central role in directing users to Wikipedia.\footnote{According to industry reports, 70-80\% of Wikipedia’s traffic originates from organic search \citep{SimilarwebWikipedia,SemRushWikipedia}.} Second, Wikimedia's public clickstream data report article-level referrals from external search engines, allowing us to measure directly the traffic channel most closely tied to AIO \citep{WikimediaClickstream}. Third, Wikipedia's multilingual structure allows us to observe traffic to different language versions of the same underlying article. This structure, together with the staggered geographic rollout of AIO, supports a within-article, difference-in-differences design.


A priori, the effect of AIO on Wikipedia traffic is ambiguous. On the one
hand, Wikipedia's concise factual content resembles the information that AIO
can synthesize and present directly, making visits to Wikipedia potentially
easy to replace. On the other hand, Wikipedia ranks prominently in organic
search and is among the sources most frequently cited in AI-generated
summaries \citep{SemrushMostCitedDomainsAI2025}. These citations could increase
Wikipedia's visibility and direct users to its pages. AIO could therefore
either substitute for Wikipedia visits by satisfying users' information needs
directly or complement them by increasing exposure to Wikipedia's content.

We construct article--language--month panels from Wikimedia's public
clickstream data covering December 2023 through December 2024. Our main samples
contain \clickMainDEPairs\ matched English--German article pairs and
\clickMainFRPairs\ matched English--French article pairs. We use English
Wikipedia as the treated edition because a substantial share of its readership
comes from the United States, where AIO became part of the default search
experience in May 2024. We use the German and French editions as controls
because their readership is concentrated in European markets where AIO did not
become part of the default search experience during our sample period
\citep{Google2024GenerativeAISearch,SearchEngineLand2025EUAIOverviews}. We then estimate a PPML difference-in-differences model that compares each English article’s monthly search traffic with that of its corresponding German or French version, controlling for article--language and calendar-month fixed effects to absorb persistent cross-language differences in baseline traffic and common monthly shocks.

We find that the introduction of AIO led to a substantial relative decline
in English Wikipedia search traffic. Monthly traffic to English articles fell
by approximately \clickMainDEDecline\ relative to their German counterparts
and by \clickMainFRDecline\ relative to their French counterparts. Translating the estimates into traffic levels, we obtain a back-of-the-envelope reduction of approximately \clickMainMonthlyLossMillions\ million direct search-originated visits to English Wikipedia per month.

We interpret these estimates as reduced-form effects of moving the U.S.\
search environment into a default-AIO regime, rather than as effects
conditional on an individual AIO display. Actual exposure was incomplete:
only a portion of English Wikipedia traffic originates in the United States,
the clickstream data combine Google with other search engines, and Google
displayed AIO for only a subset of queries while continuing to revise its
triggering rules during the post-treatment period \citep{Google2024AIOUpdate, brightedge2024AIO, seranking2024keywords}. The estimates therefore average across users, queries, search engines, and months with varying exposure. Under sign-consistent effects, this incomplete exposure would attenuate the reduced-form estimates relative to the effect among Google searches that actually displayed an AIO.

Two related papers following ours provide complimentary evidence on the consequences of AIO in different settings. \citet{agarwal2026impact} run a field experiment involving 1,000 users over a two-week period and find that AIO substantially reduces outbound organic clicks and increases zero-click searches. Their study provides direct user-level evidence over a short experimental window, while our analysis documents publisher-side traffic effects from the real-world rollout across hundreds of thousands of articles and millions of visits.\footnote{Our findings are consistent with theirs in direction, although the estimated effects in our setting are considerably smaller. This difference may reflect the nature of the publisher (Wikipedia versus generic publishers that may experience larger declines in traffic), the opt-in design of their experiment (which recruited relatively active and knowledgeable users), or differences in AIO exposure arising from the study periods and queries used.} Relatedly, \citet{zhang2026impact} study downstream activity in Reddit communities and find increases in comments and unique commenting users following AIO exposure. This positive effect on Reddit engagement contrasts with our findings, and may reflect Reddit’s distinctive position in the AIO ecosystem. Although Reddit does not typically rank among the top organic results for many searches, it is one of the sources most frequently cited and linked in AIO summaries, following its data-sharing agreement with Google \citep{edmonds2025reddit}. Taken together, these studies suggest that the effects of AIO and similar answer-first generative features can crucially depend on how it changes a publisher's visibility in the search ecosystem. Our study contributes to this emerging evidence by documenting how the rollout of AIO affected search-referral traffic to a major informational publisher, Wikipedia.


Beyond AIO within search engines, related work documents reductions in traditional search and browsing to smaller websites following standalone LLM adoption \citep{padilla2025impact}, and lower participation and question production on Stack Overflow following ChatGPT's launch \citep{burtch2024consequences,del2024large}. Within commercial platforms, generative search on Meituan changes query formulation and increases purchases \citep{ZhengEtAl2025GenerativeSearch}, while a reasoning AI assistant reduces search and bookings relative to a non-reasoning assistant on a travel platform \citep{yan2025impact}.

Focusing specifically on Wikipedia, \citet{lyu2025wikipedia} find that, following ChatGPT's November 2022 launch, recently created, popular Wikipedia articles with content more similar to ChatGPT's output experienced larger declines in viewership than less similar articles, with weaker evidence of reduced editing. Our study examines a different channel: the effect of embedding generative answers in the default interface of an incumbent search engine on referrals to publishers. We compare search-originated traffic to matched language versions of the same Wikipedia articles to assess whether a platform that traditionally directed users to publisher content begins to substitute for those visits when it supplies answers itself.

More broadly, our findings show how generative AI can change the role
of digital intermediaries. When an intermediary moves from organizing links to producing answers, it may lower users' search costs while reducing visits to
the upstream sources on which those answers rely. The results therefore have implications not only for search-engine design, but also for attribution, platform governance, and the long-run sustainability of the content ecosystem that generative intermediaries depend on.

\section{Setting and Data}
\label{sec:setting_data}

\paragraph{AIO rollout and analysis period.}
Google's rollout of AI Overviews was staggered across countries and regions. First, Google  announced nationwide U.S. availability of AIO on May~15,~2024
\citep{Google2024GenerativeAISearch}. At this time, in the U.S.\, AIO was introduced
directly into the default search experience, without requiring users to opt in.
In contrast, AIO was not part of the default search experience in the European
markets corresponding to our control-language editions during our analysis
period; Google began rolling out AIO by default in European Union regions only in early 2025
\citep{SearchEngineLand2025EUAIOverviews}.

This geographic staggering motivates our difference-in-differences design. We
treat May 2024 as the onset of the U.S. availability regime. The estimation
panel contains five pre-treatment months from December 2023 through April 2024
and eight post-treatment months from May through December 2024.

\paragraph{Treatment and control languages.}
The Wikimedia clickstream data report search traffic by language edition and
article, but not by the reader's country. We therefore proxy geographic
exposure to AIO using Wikipedia language editions. We designate the
English-language edition (\texttt{en.wikipedia.org}) as the treated group
because approximately $40\%$ of its traffic originates in the United States,
where AIO became part of the default search experience during our treatment
period. For the same underlying articles, we use the German- and
French-language editions (\texttt{de.wikipedia.org} and
\texttt{fr.wikipedia.org}) as control groups. These editions are among the
largest Wikipedia language editions by page views \citep{WikipediaTopLanguages},
and their readership is concentrated in European markets where AIO was not part
of the default search experience during our estimation window.\footnote{According
to \cite{Similarweb}, approximately $77\%$ of German Wikipedia traffic comes
from Germany, $8\%$ from Austria, and $5\%$ from Switzerland, while approximately
$77\%$ of French Wikipedia traffic comes from France, $5\%$ from Belgium, and
$5\%$ from Canada.} Moreover, prior
research documents meaningful similarities in online information discovery and
consumption behaviors across users in the United States, Germany, and France
\citep{stier2020populist,miz2020trending,lemmerich2019world,fletcher2018automated},
supporting the use of German and French editions as comparison groups. 

Spanish provides a less suitable comparison because it is
widely used across multiple countries, including the United States, making its
language edition a less precise proxy for geographic exposure. We also exclude
Russian and Chinese editions due to censoring concerns. As an additional
robustness check, we report English--Japanese comparisons separately in Web
Appendix~$\S$\ref{appsec:japanese_control}.

\paragraph{Search referrals traffic data.}
We obtain article-level clickstream data from Wikimedia's publicly available data sets \citep{WikimediaClickstream}. Among the available traffic sources, we use \texttt{other-search}, which records monthly human referrals from external
search engines as our traffic metric. Because Google accounts for approximately $90\%$ of the global search market \citep{statistics2026google}, this measure provides a close proxy for Google search referrals to Wikipedia. Our outcome, $R_{i,l,m}$, is therefore the monthly number of external-search referrals received by article $i$ in language edition $l$ during month $m$.\footnote{We do not use traffic from
other websites as an untreated comparison outcome because AIO may also affect traffic to those websites and, consequently, their outbound referrals to
Wikipedia. Non-search referrals therefore cannot be treated as a clean control outcome.}

\paragraph{Article selection and panel construction.}
We construct separate English--German and English--French article samples by
matching the same underlying Wikipedia articles across language editions.
Wikimedia does not report monthly \texttt{other-search} counts below 10; these
observations are therefore censored in the clickstream data. To ensure that
our main analysis is based on reliably measured search-referral counts, we
retain articles that exist since November 2023, and for which both editions have a reported \texttt{other-search}
count of at least 10 in every pre-treatment month (December 2023 through April
2024). This restriction ensures that sample membership is determined using only
pre-treatment information and gives us \clickMainDEPairs\ and
\clickMainFRPairs\ matched articles in the English--German and English--French
samples, respectively.

For post-treatment observations that are censored by Wikimedia's reporting
threshold, we assign a value of 10 to English counts and a value of 0 to
control-language counts. This directional assignment is conservative for
detecting a negative English effect because it reduces the measured decline in
English search referrals relative to the control editions.\footnote{Alternative
assignments of censored post-treatment observations do not meaningfully affect
the results. Specifically, reversing the 0 and 10 assignment, assigning censored observations a value of 10 in
both the English and control editions, or assigning a value of 0 in both
editions, produces estimates that are nearly identical to the baseline
results; Web Appendix~$\S$\ref{appsec:clickstream_imputation_sensitivity} for details.}

\begin{table}[htp!]
\centering
\small
\caption{Summary statistics of monthly search traffic for the English--German and English--French samples.}
\label{tab:sumstats_langs}
\setlength{\tabcolsep}{2pt}%
\makebox[\textwidth][c]{%
\begin{tabular}{lcccc}
\toprule
\midrule
\multicolumn{5}{c}{Pre-treatment months (Dec.~2023 -- Apr.~2024)} \\
\midrule
Statistic & English (EN--DE) & German & English (EN--FR) & French \\
\midrule
Mean & 4,222.13 & 624.45 & 4,305.29 & 577.76 \\
Std. & 26,615.55 & 4,061.18 & 26,872.24 & 3,842.78 \\
Min. & 10.00 & 10.00 & 10.00 & 10.00 \\
25\% & 159.00 & 38.00 & 188.00 & 36.00 \\
50\% & 628.00 & 102.00 & 721.00 & 92.00 \\
75\% & 2,438.00 & 358.00 & 2,625.00 & 312.00 \\
Max. & 12,503,258.00 & 3,061,107.00 & 12,503,258.00 & 2,386,870.00 \\
Article-Language & 499,927 & 499,927 & 530,873 & 530,873 \\
\midrule
\multicolumn{5}{c}{Post-treatment months (May~2024 -- Dec.~2024)} \\
\midrule
Statistic & English (EN--DE) & German & English (EN--FR) & French \\
\midrule
Mean & 3,559.10 & 556.75 & 3,624.94 & 511.09 \\
Std. & 21,816.49 & 3,224.70 & 22,062.84 & 4,070.44 \\
Min. & 10.00 & 0.00 & 10.00 & 0.00 \\
25\% & 136.00 & 32.00 & 159.00 & 28.00 \\
50\% & 541.00 & 89.00 & 614.00 & 75.00 \\
75\% & 2,101.00 & 315.00 & 2,247.00 & 262.00 \\
Max. & 11,251,661.00 & 1,242,785.00 & 11,251,661.00 & 2,286,315.00 \\
Article-Language & 499,927 & 499,927 & 530,873 & 530,873 \\
\bottomrule
\end{tabular}
}
\end{table}

\paragraph{Summary Statistics}
\enlargethispage{\baselineskip}

Table~\ref{tab:sumstats_langs} reports summary statistics of monthly search
traffic for each sample, separately for the pre-treatment period (December
2023 to April 2024) and the post-treatment period (May 2024 to December 2024).
Because we independently construct the English--German and English--French
samples, we summarize English search traffic separately for the two comparison
samples. We observe three broad patterns in the data. First, English Wikipedia
articles receive substantially more search traffic than their German and French
counterparts. As we discuss in the empirical analysis section, this difference
in scale motivates our use of PPML rather than a log-linear specification.
Second, search traffic is highly right-skewed across languages: the mean exceeds
the median, reflecting a small number of high-traffic articles. Third, average
monthly search traffic declines from the pre- to post-treatment period across
all language editions. In the empirical analysis below, we examine whether
English search traffic declined more than the corresponding control editions
following the AIO rollout.

\section{Empirical Analysis}
\label{sec:empirical_analysis}

\subsection{Model-Free Analysis}
\label{ssec:model_free}

We now present a model-free comparison of the English and control-language traffic paths before and after the introduction of Google AIO in the United States.

\begin{figure}[htp!]
\centering
\begingroup
\small
\setlength{\tabcolsep}{2pt}
\begin{tabular}{@{}>{\centering\arraybackslash}m{0.49\textwidth}>{\centering\arraybackslash}m{0.49\textwidth}@{}}
English--German & English--French \\
\paperassetgraphic[width=\linewidth,keepaspectratio]{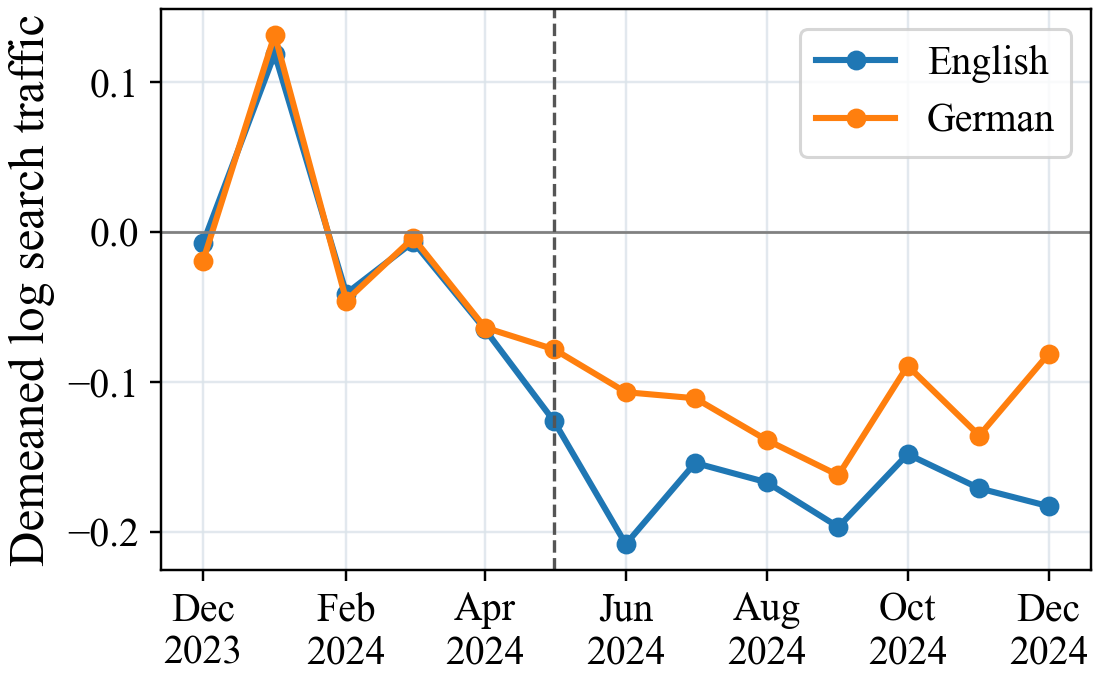} & \paperassetgraphic[width=\linewidth,keepaspectratio]{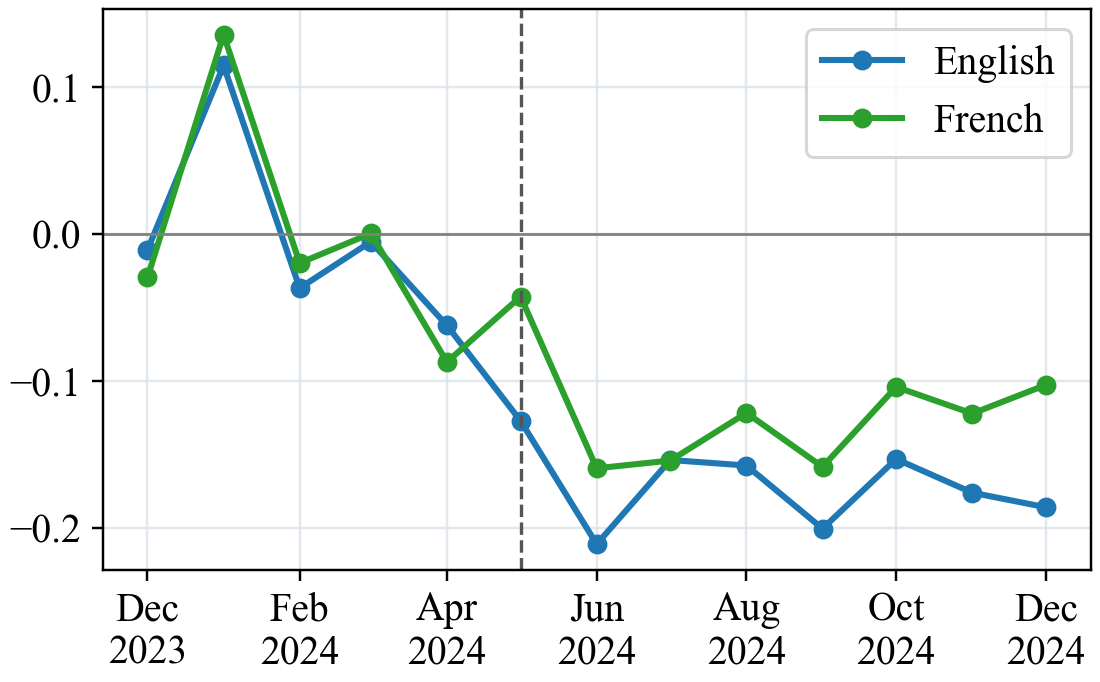} \\
\end{tabular}
\endgroup
\caption{Demeaned log total monthly search traffic through December 2024 for the English--German and English--French samples. Each series is demeaned using December 2023--April 2024; the dashed line marks May 2024. The English--German and English--French panels contain 499,927 and 530,873 matched article pairs, respectively, with 13 monthly observations in each plotted series.}
\label{fig:clickstream_model_free}
\end{figure}

Figure~\ref{fig:clickstream_model_free} plots these comparisons over the entire analysis period. For each language and month, we sum $R_{i,l,m}$ across all the
articles in corresponding sample, transform the aggregate using $\log(1+R)$, and demean each series by its own pre-treatment (December
2023--April 2024) average. The log transformation emphasizes proportional
rather than absolute changes, while demeaning places the English and control
series on a common pre-treatment baseline.

Before May 2024, the English and control-language series move closely
together. After the AIO rollout, English search traffic shifts downward
relative to both German and French and generally remains below the control
series through the end of the sample. These patterns provide initial evidence
of a negative post-AIO change in English search traffic, which we estimate
formally in the next subsection using a PPML difference-in-differences model.
Because the figure aggregates traffic across articles and applies a log
transformation, it does not correspond one-for-one to the article-level PPML
estimand. Nevertheless, both approaches characterize proportional changes, so
the figure provides a useful descriptive preview of the regression results.

\subsection{PPML Difference-in-Differences Specification and Main Results}
\label{ssec:pooled_ppml_did}

We next quantify the post-AIO divergence shown in
Figure~\ref{fig:clickstream_model_free} using a PPML
difference-in-differences model at the article--language--month level.

We use PPML rather than an OLS difference-in-differences model with logged
search traffic for three reasons. First, when the treated and control outcome
distributions differ substantially in scale, a log-linear specification can
produce a treatment effect with the opposite sign from the aggregate effect of
interest \citep{mcconnell2024can}. This concern is especially relevant in our
setting because English articles receive substantially more traffic than their
German and French counterparts. Second, OLS on logged traffic gives similar
weight to a proportional change on a low-traffic article and the same
proportional change on a high-traffic article, so its estimand more closely
reflects the response of a typical article. Our primary interest instead lies
in total traffic to Wikipedia, for which changes on high-traffic articles
should contribute more. By modeling the conditional mean of traffic directly
in levels, PPML aligns more closely with this aggregate-traffic estimand while
retaining a multiplicative, proportional interpretation
\citep{wooldridge2010econometric}. Third, PPML accommodates observations with
zero traffic directly, without requiring us to drop them or apply a
transformation such as $\log(1+R)$.

We estimate the following PPML difference-in-differences
specification separately for each control language
$c\in\{\mathrm{DE},\mathrm{FR}\}$:
\begin{equation}
\label{eq:pooled_ppml_did}
\mathbb{E}\!\left[R_{i,l,m}\mid \alpha_{i,l},\delta_m\right]
=
\exp\!\left(
\alpha_{i,l}
+
\delta_m
+
\beta_c\big(\text{English}_l \times \text{Post}_m\big)
\right),
\end{equation}
where $i$ indexes the underlying article, $l$ indexes the language edition,
and $m$ indexes the month. The indicator $\text{English}_l$ equals one for the
treated English edition, and $\text{Post}_m$ equals one beginning in May 2024,
the month in which AIO became part of the default U.S.\ search experience.
Article$\times$language fixed effects, $\alpha_{i,l}$, absorb persistent
differences in article popularity and audience size across language editions,
while month fixed effects, $\delta_m$, absorb temporal shocks common to both
editions in a comparison.

Under the parallel-trends assumption, $\beta_c$ identifies the average
post-AIO change in English search traffic relative to control language $c$.
This assumption requires that, absent AIO, search traffic to the English and
control-language versions of the same articles would have continued to evolve
proportionally in parallel. The coefficient's exact percentage interpretation
is $100\times[\exp(\beta_c)-1]$. We assess the parallel-trends assumption and
the sensitivity of our conclusions to deviations from it in
Section~\ref{ssec:clickstream_robustness} and Web
Appendix~$\S$\ref{appsec:clickstream_parallel_pretrends}. Further, we two-way cluster standard errors by article and month. Clustering by article
allows for correlated shocks across language editions and over time within an
article, while clustering by month allows for common shocks across articles in
a given month \citep{Bertrand2004,Cameron2011}.

\begin{table}[htp!]
\centering
\small
\setlength{\tabcolsep}{2.0pt}
\begin{threeparttable}
\caption{PPML DiD estimates}
\label{tab:clickstream_ppml_did}
\begin{tabular}{lcc}
\toprule
 & English--German & English--French \\
\midrule
English $\times$ post & $-0.0561^{***}$ & $-0.0494^{***}$ \\
 & (0.0117) & (0.0136) \\[0.25em]
No. observations & 12,998,102 & 13,802,698 \\
Articles (QIDs) & 499,927 & 530,873 \\
\midrule
Article $\times$ language FE & Yes & Yes \\
Month FE & Yes & Yes \\
\bottomrule
\end{tabular}
\begin{tablenotes}[flushleft]\small
\item Robust standard errors in parentheses are two-way clustered by article and month.
\item Significance: $^{***}p<0.001$, $^{**}p<0.01$, $^{*}p<0.05$.
\end{tablenotes}
\end{threeparttable}
\end{table}

Table~\ref{tab:clickstream_ppml_did} reports the estimates. The English--German coefficient of
$\clickMainDEBeta$ implies that monthly English Wikipedia search traffic
declined by approximately $\clickMainDEDecline$ relative to German following
the AIO rollout. The English--French coefficient of $\clickMainFRBeta$ implies
a corresponding decline of approximately $\clickMainFRDecline$ relative to
French. The similarity in the signs and magnitudes of the estimates across two separately constructed control samples reduces the likelihood that a shock unique to either control edition drives the results.

We interpret these coefficients as reduced-form effects of making AIO part of
the default U.S.\ search experience, analogous to intent-to-treat effects. They do not identify the effect of an individual AIO display because realized
exposure varied along several dimensions, e.g., only a portion of English Wikipedia traffic originates in the United States, and Google displayed AIO for only a subset of queries \citep{Google2024GenerativeAISearch}. Google also
continued to revise AIO's triggering rules and technical implementation after
the May 2024 rollout, changing both the prevalence of AIO and the composition
of queries for which it appeared \citep{Google2024AIOUpdate, brightedge2024AIO, seranking2024keywords}.\footnote{Third-party audits suggest that AIOs now appear on approximately 18\% of Google searches \citep{Tabuena2025OptimizeAIO, Pew2025AISummariesClicks}, though this number varies by query type and measurement timeframe \citep{aral2026rise}.} Our estimates therefore average across users, queries, and months with different levels of realized AIO exposure. Because we do not observe query-level AIO displays, we cannot recover the effect conditional on an AIO display. Under sign-consistent effects, the inclusion of unexposed traffic likely attenuates the reduced-form estimates relative to the effect among Google searches that actually displayed an AIO.

\subsection{Robustness Checks}
\label{ssec:clickstream_robustness}

We next evaluate the robustness of our main estimates by examining  whether alternative data choices, treatment timing, or control groups
change our conclusions. We first assess the parallel-trends assumption using complementary diagnostics and then summarize the remaining robustness checks; the corresponding Web Appendix reports the full specifications, tables, and figures.

\squishlist

\item \textbf{Parallel trends.} A key assumption of the DiD model is that, absent AIO, English search traffic would have followed the same trend as control-language search traffic \citep{AngristPischke2009}. We assess this assumption in four ways and confirm that there are no significant violations that threaten the main takeaways. First, a parametric linear pre-treatment test finds no statistically significant pre-treatment drift. Second, monthly event studies show some pre-treatment differences, but the German estimates fluctuate around zero, the French estimates move opposite the post-AIO decline, and both are small relative to the post-treatment estimates. Next, we use Honest DiD to examine how large departures from parallel trends must be to overturn the result; the confidence intervals remain below zero when post-treatment changes are allowed to reach 40\% of the largest pre-treatment change for German and 35\% for French. Finally, both equal-weighted and precision-weighted linear extrapolations of the pre-treatment event-study coefficients place observed post-treatment search traffic significantly below its counterfactual for both controls. See Web Appendix~$\S$\ref{appsec:clickstream_parallel_pretrends} for the detailed analyses.

\item \textbf{Sensitivity to post-treatment censoring assignment.}  Wikimedia suppresses monthly article-level search counts below 10. In the main analysis, we address this missing data issue by assigning unpublished post-treatment counts a value of 10 for English and 0 for the control language, which works against finding a negative English effect. We test whether the findings depend on this treatment by reversing the assignment---0 for English and 10 for the control---while leaving all published counts and the estimation samples unchanged. Under this opposite assignment, the estimates imply a 5.5\% decline in English search traffic relative to German and a 4.9\% decline relative to French; both estimates remain statistically significant and are nearly identical to the main results. Thus, the estimated effect is not driven by how we assign censored post-treatment observations. Please see Web Appendix~$\S$\ref{appsec:clickstream_imputation_sensitivity} for the detailed analysis.

\item \textbf{Pre-treatment timing placebo.} We also run placebo tests to check whether the main estimates reflect a relative decline in English traffic that began before AIO rather than a break following its rollout. Accordingly, we re-estimate the PPML DiD model using only pre-treatment period's data (December 2023--April 2024) and designate March--April 2024 as an artificial post-treatment period. The placebo estimates are not statistically significant for either German or French. The absence of a negative English-relative-control shift before May supports the interpretation that the main estimates capture a post-AIO divergence. Please see Web Appendix~$\S$\ref{appsec:clickstream_preperiod_timing_placebo} for details.

\item \textbf{Alternative control language.}
As an additional comparison, we construct a sample of
\clickMainJAPairs\ matched English--Japanese article pairs and end the analysis
before August 2024, when Google expanded AIO to Japan \citep{Google2024AIOAugExpansion}. The PPML estimate implies a \clickMainJADecline\ decline in English search
traffic relative to Japanese. Because the Japanese edition differs substantially from the main
controls in audience composition and because this comparison uses a shorter
post-treatment window, we treat it as corroborating evidence on the direction
of the effect rather than pool it with the German and French estimates. Please see Web
Appendix~$\S$\ref{appsec:japanese_control} for details.

\squishend

\section{Discussion and Implications}
\label{sec:discussion_implications}

We now place our empirical findings in a broader economic context. We first discuss the magnitude and scope of the measured traffic effect, and then consider the implications for publisher monetization, content supply, intermediary design, welfare, and policy.

\paragraph{Economic magnitude.}
We first translate the proportional estimates into traffic levels to illustrate their economic scale. We first take the arithmetic mean of the
English--German and English--French PPML coefficients, which corresponds to an
average decline of approximately \clickMainEconomicDecline. We then apply this
average effect to the arithmetic mean of monthly post-treatment English search
traffic in the two comparison samples. Because the observed post-treatment
traffic already reflects the estimated decline, we use the multiplicative
structure of the PPML model to recover the corresponding counterfactual traffic
level absent AIO. This calculation implies approximately
\clickMainMonthlyLossMillions\ million fewer direct search-originated visits
to English Wikipedia per month. If the estimated monthly effect persisted for
one year, it would correspond to approximately
\clickMainAnnualLossBillions\ billion fewer direct search visits. Web
Appendix~$\S$\ref{appsec:economic_magnitude} reports the complete calculation
and assumptions.

\paragraph{Scope of the estimates.}
We next clarify what our estimates capture. Our outcome measures visits that arrive directly from external search engines. We therefore interpret the estimates as the first-order effect of AIO on search-originated traffic to Wikipedia, rather than as the overall effect on Wikipedia traffic across all referral channels.

AIO may generate additional downstream effects within Wikipedia. For example, a user who enters Wikipedia through a search
engine may visit an initial article and then follow links to several other
Wikipedia pages.\footnote{Internal article-to-article referrals account for roughly 40\% of the overall Wikipedia traffic \citep{piccardi2023large}.} If AIO eliminates the initial search visit, Wikipedia may lose
both the entry-page view and the internal navigation that would have followed
it. The public clickstream data identify only the immediate referrer for each
article transition; they do not link an internal click to the external source
that initiated the user's session \citep{WikimediaClickstream}. We therefore
cannot determine how much internal traffic originates from search-referred
sessions or quantify the resulting downstream effect.
Accordingly, our estimates capture the direct effect of AIO on traffic referred by external search engines. The total effect on Wikipedia traffic could be larger if lost search visits also result in fewer internal referrals.


\paragraph{Publisher monetization and content supply.}
Wikipedia does not monetize its traffic through advertising or subscriptions,
but many publishers rely directly on search-originated visits for advertising
impressions, subscription acquisition, lead generation, and commerce. For
these publishers, a decline in search traffic reduces both immediate monetization opportunities and the number of users entering the top of the conversion funnel.

To illustrate the potential advertising implications, we assume that each
foregone search visit corresponds to one foregone landing-page view and apply
a page-revenue range of approximately \$8.99--\$30.81 per thousand page views.
We derive this range using an average of 5.22 display-ad opportunities per
page, fill rates of 75--90\%, a 70\% publisher revenue share, and CPMs of
\$3.28--\$9.37
\citep{ZengKohnoRoesner2020BadNews,GoogleAdManagerFillRate,
Playwire2026StatePublisherAdRevenue,ANA2025ProgrammaticBenchmarkQ1}.
Applying these benchmarks to the estimated traffic reduction yields an
illustrative advertising-revenue equivalent of approximately
\$\clickMainMonthlyRevenueLowMillions M--\$\clickMainMonthlyRevenueHighMillions M
per month, or
\$\clickMainAnnualRevenueLowMillions M--\$\clickMainAnnualRevenueHighMillions M
per year if the estimated monthly effect persisted. These figures illustrate the potential monetization consequences for an ad-supported
publisher experiencing a traffic decline of the estimated magnitude. Web
Appendix~$\S$\ref{appsec:economic_magnitude} reports the complete calculation
and assumptions.

Traffic displacement may also affect the long-run supply of content even when
a publisher does not monetize visits directly. Prior research on Wikipedia suggests that greater audience reach and article popularity can influence contributors' incentives to produce and maintain content \citep{zhang2011group,champion2025countering}. By reducing the
audience reaching source pages, answer engines may weaken the visibility,
recognition, and other benefits that motivate content production. This creates
a potential feedback loop: search platforms use publisher content to generate
answers, but those answers may reduce the traffic and incentives that support
the production and maintenance of that content. Our analysis does not estimate
this longer-run response, but the traffic displacement that we document
identifies an important first step in that mechanism.

\paragraph{Intermediary design and welfare.}
Our findings highlight a fundamental difference between link-based aggregation
and answer-producing intermediation. Traditional search engines and news
aggregators primarily lower discovery costs and route users to upstream
content. Consistent with this mechanism, prior research often finds that
link-based aggregation complements publisher consumption
\citep{chiou2017content,calzada2020news,athey2021impact}. AIO changes the
intermediary's role by synthesizing the underlying information and presenting
an answer before the organic links. By satisfying some users' information
needs directly, the intermediary can substitute for a visit to the source
rather than merely facilitate it.

The negative effects that we estimate are particularly notable because
Wikipedia is among the sources most frequently cited in AI-generated summaries
\citep{SemrushMostCitedDomainsAI2025}. These citations may direct some users
to Wikipedia, but our estimates indicate that any resulting traffic did not
offset the average decline associated with the introduction of AIO. More
generally, the results suggest that an intermediary can shift from
complementing upstream content to substituting for source-page visits when it
moves from organizing links to producing answers.

The observed traffic displacement does not, by itself, imply that AIO reduces
overall welfare. Answer-first search may lower users' search costs, provide
faster access to information, and direct users who do click to more relevant
pages
\citep{Google2024GenerativeAISearch,Google2025AISearchClicks}. Our design
measures publisher traffic rather than consumer surplus, answer quality, or
the value of users' time. The results thus highlight an ecosystem level tradeoff: the same interface that improves the immediate efficiency of
information retrieval may shift attention away from the publishers that
produce and maintain the underlying information. The long-run welfare effects
therefore depend not only on users' short-run benefits but also on how reduced
traffic changes publishers' incentives to supply high-quality content.

\paragraph{Platform and policy implications.}
Our findings also raise broader questions about market power, attribution, and
the governance of AI-mediated information markets. Search platforms control
the interface through which users discover content and can reallocate traffic
through product-design changes that publishers cannot directly control. As more discovery and consumption take place within search and AI platforms, these platforms also retain a larger share of user-interaction data, measurement
feedback, and monetization opportunities. These changes can deepen publishers'
dependence on platform-controlled ranking, advertising, and attribution
systems, reinforcing longstanding concerns about digital gatekeepers
\citep{greenstein2010gatekeeping, goldfarb2019digital}.

Our results highlight the value of greater transparency about AIO exposure,
source attribution, and downstream traffic effects. They also motivate the
design of mechanisms that preserve incentives to produce content while
allowing users to benefit from AI-generated summaries. Potential approaches
include effective publisher controls, auditable traffic reporting,
compensation or revenue-sharing arrangements, and attribution systems that
value individual sources' contributions to generated answers, e.g., through Shapley-based attribution systems
\citep{YeYoganarasimhan2025}. Recent regulatory debates over publisher
opt-outs and litigation concerning the use of publisher content in AI
summaries reflect these broader tensions
\citep{Digiday2026AIOOptOut,Reuters2026GoogleAISummariesLawsuit}. As search
platforms increasingly produce answers rather than merely rank links, the
design of these mechanisms will become central to sustaining both efficient
information access and the upstream content on which it depends.

\section{Conclusion}
\label{sec:conclusion}

Generative AI is changing search engines from intermediaries that organize
links into systems that answer users' questions within the search interface.
We study the publisher-side consequences of this shift by combining Google's
staggered rollout of AI Overviews with Wikimedia's article-level clickstream
data. Comparing English Wikipedia articles with the same articles in German
and French, we find that default AIO availability reduced monthly English
search traffic by approximately \clickMainDEDecline\ relative to German and
\clickMainFRDecline\ relative to French. 

Our findings imply that generative-answer features such as AIO in search engines can materially reallocate attention away from publisher pages, even when it cites and links to that content. Although this shift may lower users' search costs and improve access to information, it can also reduce publishers' monetization opportunities and weaken the incentives that sustain
the underlying content ecosystem. As digital intermediaries increasingly
produce answers rather than merely organize links, preserving the incentives
to create and maintain the content on which those answers rely will become
central to the economics of AI-mediated information.

\section*{Funding and Competing Interests Declaration}
Author(s) have no competing interests to declare.

\clearpage
\putbib
\end{bibunit}
\singlespacing

\newpage
\begin{appendices}

\setcounter{table}{0}
\setcounter{figure}{0}
\setcounter{equation}{0}
\setcounter{page}{0}
\renewcommand{\thetable}{A\arabic{table}}
\renewcommand{\thefigure}{A\arabic{figure}}
\renewcommand{\theequation}{A\arabic{equation}}
\renewcommand{\thepage}{\roman{page}}
\pagenumbering{roman}
\begin{bibunit}

\section{Search Results Pages with and without AIO}
\label{appsec:serps}
\begin{figure}[htp!]
  \centering
  \paperassetgraphic[width=0.65\linewidth]{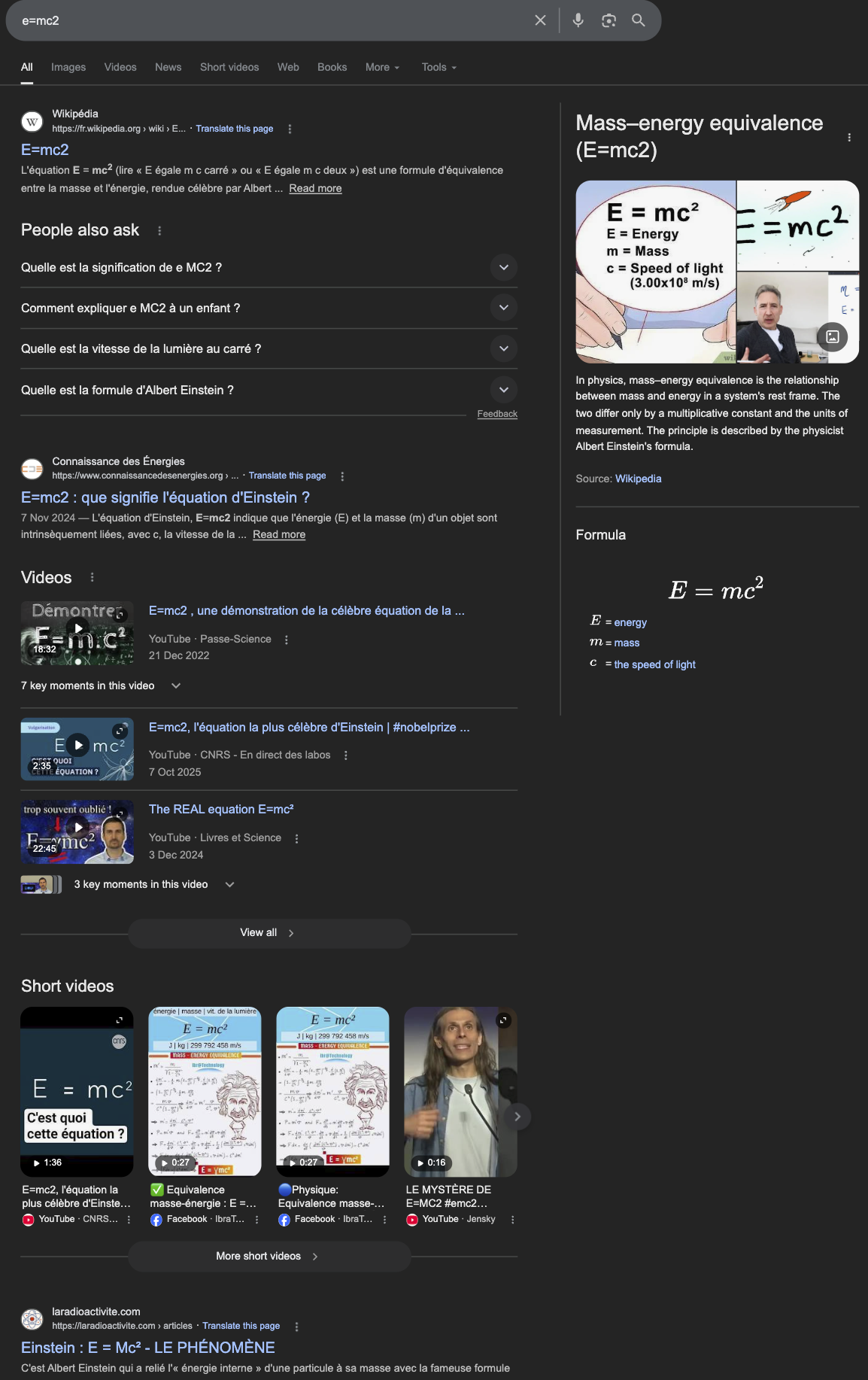}
  \caption{A Google search result page {\it without} AIO for the query ``e=mc2.''}
  \label{fig:serp_wo_aio}
\end{figure}

\begin{figure}[htp!]
  \centering
  \paperassetgraphic[width=0.65\linewidth]{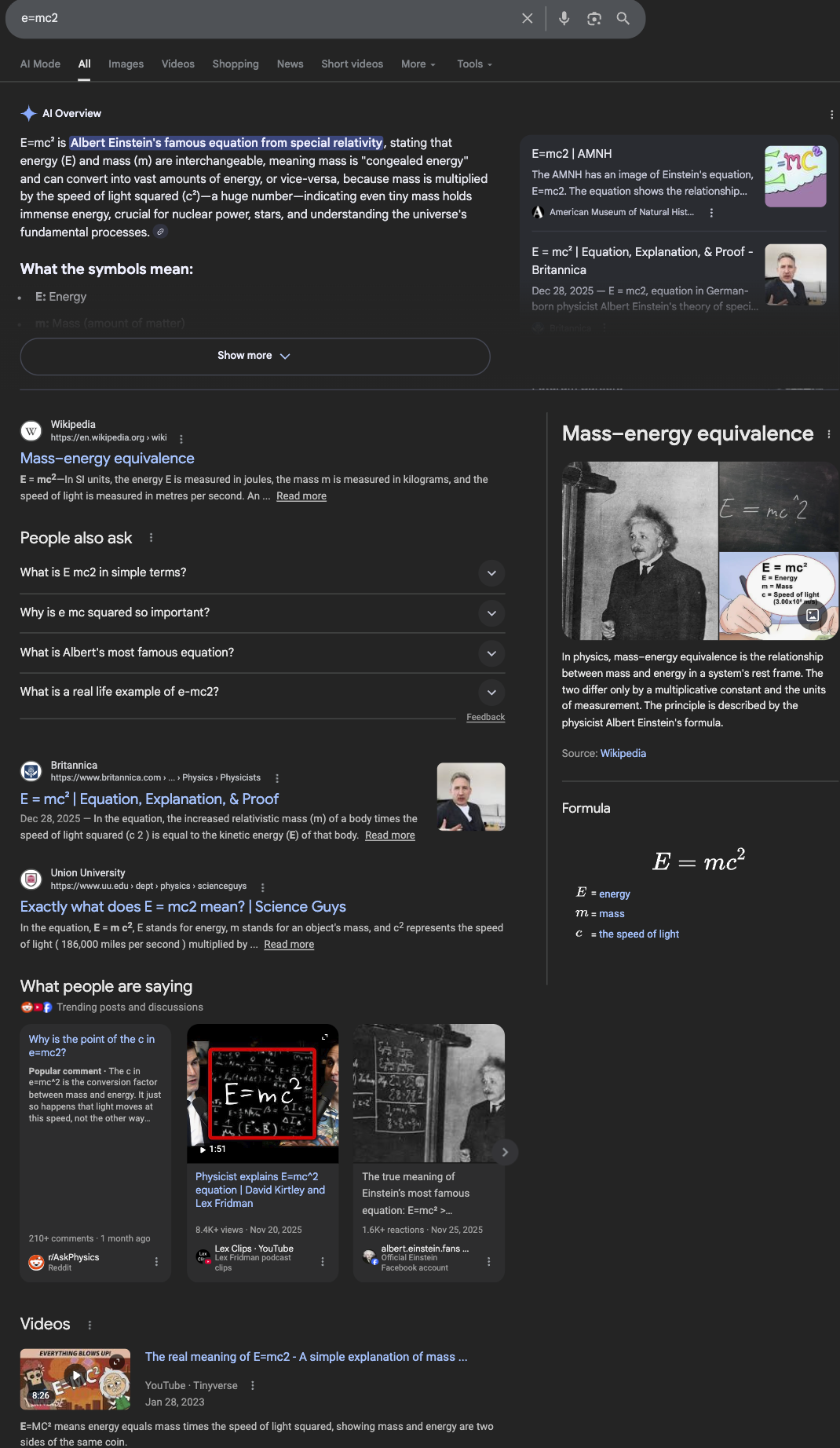}
  \caption{A Google search result page {\it with} AIO for the query ``e=mc2.''}
  \label{fig:serp_w_aio}
\end{figure}

In this section, we present two examples of Google's search engine results (SERPs) without and with AIO. As we can see in Figures \ref{fig:serp_wo_aio} and \ref{fig:serp_w_aio}, AIO pushes the top search results (including Wikipedia in this example) down in the SERP.
\pagebreak

\clearpage

\section{Parallel Pre-Treatment Trends}
\label{appsec:clickstream_parallel_pretrends}
The main assumption in our difference-in-difference specification is that the treatment and control articles would have parallel trends absent the AIO treatment. In this section, we examine this {\it parallel-trends} assumption for the pairwise samples of English vs. German and French control languages using linear pre-treatment tests, dynamic event studies, Honest DiD sensitivity analysis, and linear counterfactual extrapolations.

\subsection{Linear Pre-Treatment Tests}
\label{appsec:clickstream_linear_pretrend_tests}
As a simple parametric check, we estimate a PPML differential-trend model using the set of five pre-treatment months, denoted as $\mathcal{M}_{\mathrm{pre}}$, for estimation, December 2023--April 2024. For each sample, we estimate:
\begin{equation}
\label{eq:pooled_ppml_pretrend}
\mathbb{E}\!\left[R_{i,l,m}\mid \alpha_{i,l},\delta_m\right]
=
\exp\!\left(
\alpha_{i,l}
+
\delta_m
+
\eta_c\big(\text{English}_l \times t_m\big)
\right),
\qquad m\in\mathcal{M}_{\mathrm{pre}}.
\end{equation}
Here, $t_m$ is a linear month index within the pre-treatment estimation period. Similar to our main two-way fixed-effects (TWFE) specification, $\alpha_{i,l}$ and $\delta_m$ denote article-language and month fixed effects, respectively. For control $c$, the coefficient $\eta_c$ captures whether English search traffic was already drifting linearly relative to that control-language edition before treatment. Under the parallel-trends benchmark, we expect $\eta_c\approx 0$. Table~\ref{tab:clickstream_ppml_pretrend} reports the results of this test.

\begin{table}[htp!]
\centering
\small
\setlength{\tabcolsep}{2.0pt}
\begin{threeparttable}
\caption{PPML linear pre-treatment trend tests}
\label{tab:clickstream_ppml_pretrend}
\begin{tabular}{lcc}
\toprule
 & English--German & English--French \\
\midrule
English $\times t_m$ & $-0.0014$ & $0.0028$ \\
 & (0.0035) & (0.0076) \\[0.25em]
No. observations & 4,999,270 & 5,308,730 \\
Articles (QIDs) & 499,927 & 530,873 \\
\midrule
Article $\times$ language FE & Yes & Yes \\
Month FE & Yes & Yes \\
\bottomrule
\end{tabular}
\begin{tablenotes}[flushleft]\small
\item Robust standard errors in parentheses are two-way clustered by article and month.
\item Significance: $^{***}p<0.001$, $^{**}p<0.01$, $^{*}p<0.05$.
\end{tablenotes}
\end{threeparttable}
\end{table}

The estimated monthly differential slopes are $-0.0014$ for German and $0.0028$ for French. Neither is statistically different from zero under the two-way-clustered inference. We next use the event-study specification to assess dynamic deviations that a single linear slope can miss.

\subsection{PPML Event Studies}
\label{ssec:pooled_ppml_event_study_results}
\begin{figure}[htp!]
\centering
\begingroup
\small
\setlength{\tabcolsep}{2pt}
\begin{tabular}{@{}>{\centering\arraybackslash}m{0.49\textwidth}>{\centering\arraybackslash}m{0.49\textwidth}@{}}
English--German & English--French \\
\paperassetgraphic[width=\linewidth,keepaspectratio]{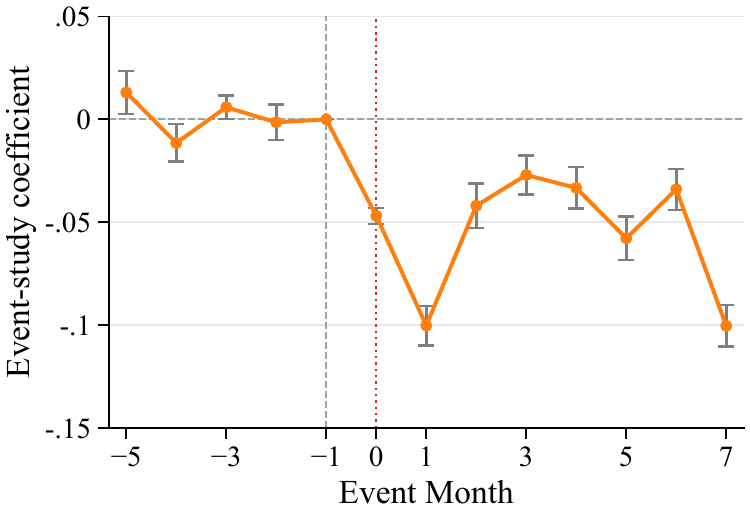} & \paperassetgraphic[width=\linewidth,keepaspectratio]{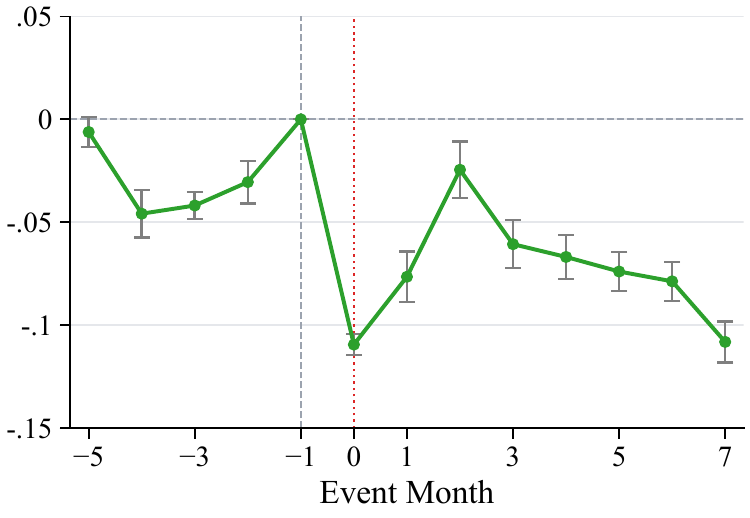} \\
\end{tabular}
\endgroup
\caption{Monthly PPML event-study estimates through December 2024 for the English--German and English--French samples. April 2024 is the omitted reference month and May 2024 is event month zero. Pointwise 95\% confidence intervals use the direct two-way-clustered standard errors and asymptotic-normal inference from Stata.}
\label{fig:clickstream_event_studies}
\end{figure}

The DiD specification summarizes the post-treatment period with a single average effect. We also estimate a pooled event-study that replaces $\text{Post}_m$ with event-month interactions:
\begin{equation}
\label{eq:pooled_ppml_event_study}
\mathbb{E}\!\left[R_{i,l,m}\mid \alpha_{i,l},\delta_m\right]
=
\exp\!\left(
\alpha_{i,l}
+
\delta_m
+
\sum_{r\in\mathcal{M}\setminus\{m_0\}}
\beta_{c,r}
\big(\text{English}_l \times \mathbf{1}\{m=r\}\big)
\right),
\end{equation}
where $\mathcal{M}$ is the set of event months and $m_0=-1$ is the omitted reference month  (April 2024), and May 2024 is the event month. For control $c$, each $\beta_{c,r}$ measures the English-versus-control difference in event month $r$ relative to April. Figure~\ref{fig:clickstream_event_studies} reports the German and French estimates through December 2024.

Figure~\ref{fig:clickstream_event_studies} shows pre-treatment fluctuations that the linear tests do not capture. Although some of the pre-treatment coefficients have 95\% confidence intervals that do not include zero, three observations make the violations less concerning. In the English-German sample, the violations are more fluctuations around zero, which do not represent a negative pre-treatment trend. In the English-French sample, the violations exhibit a pre-treatment positive trend that runs counter to our estimated negative treatment effect. This makes our estimation a lower bound for the effect's magnitude. Finally, the magnitude of the pre-treatment violations is much smaller than the magnitude of the post-treatment coefficient, making the overall results less sensitive to these deviations from the parallel pre-treatment trends. Nevertheless, we use Honest DiD in the next section to report how sensitive the average post-treatment conclusion is to deviations calibrated from the observed pre-period.

\subsection{Honest DiD Sensitivity under Relative-Magnitude Restrictions}
\label{appsec:clickstream_honest_did}

As we discussed in the previous section, several pre-treatment event-study coefficients are close to, but statistically different from, zero. We therefore use the Honest DiD framework of \citet{rambachan2023more} to examine the sensitivity of our conclusions to violations of the parallel-trends assumption. Following their notation, we decompose each sample event-study coefficient as
\begin{equation}
\beta_{c,r}
=
\tau_{c,r}
+
\delta_{c,r},
\end{equation}
where $\tau_{c,r}$ is the causal effect in event month $r$ and $\delta_{c,r}$ is the differential trend between English and control language $c$ that would have occurred absent treatment. Under no anticipation, $\tau_{c,r}=0$ in the pre-treatment period, so the pre-treatment coefficients identify the corresponding differential trends. The omitted reference month, April, 2024 here, is $m_0=-1$, with $\delta_{c,m_0}=0$ by normalization.

\begin{figure}[htp!]
\centering
\begingroup
\small
\setlength{\tabcolsep}{2pt}
\begin{tabular}{@{}>{\centering\arraybackslash}m{0.49\textwidth}>{\centering\arraybackslash}m{0.49\textwidth}@{}}
English--German & English--French \\
\paperassetgraphic[width=\linewidth,keepaspectratio]{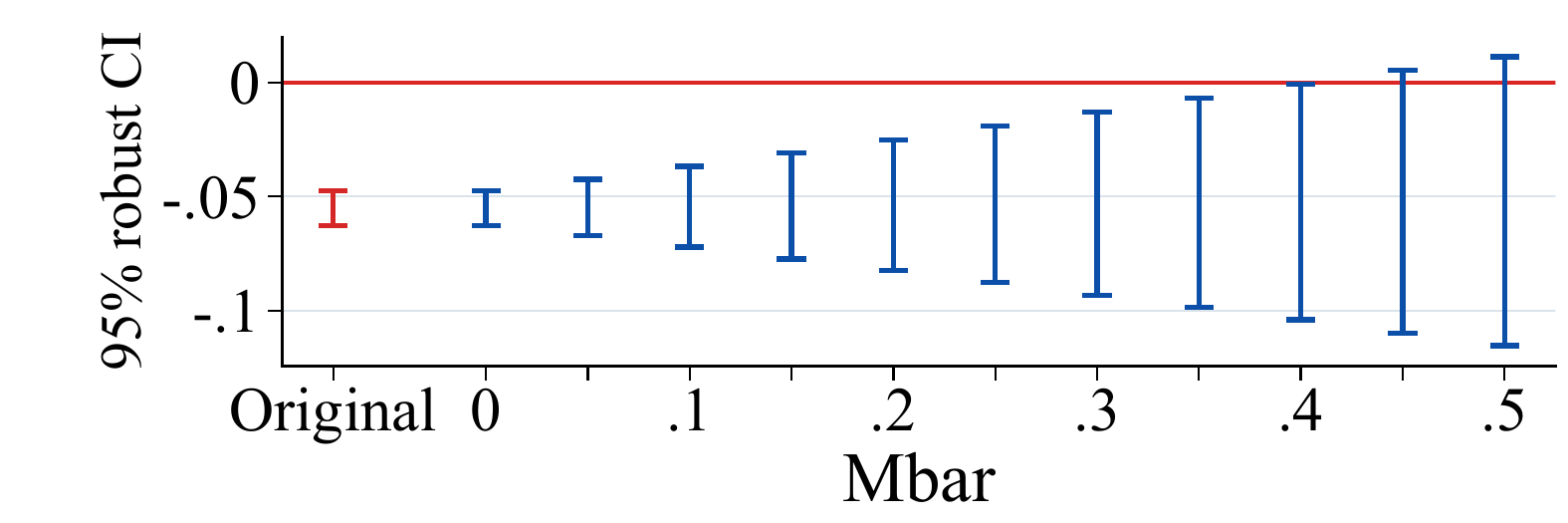} & \paperassetgraphic[width=\linewidth,keepaspectratio]{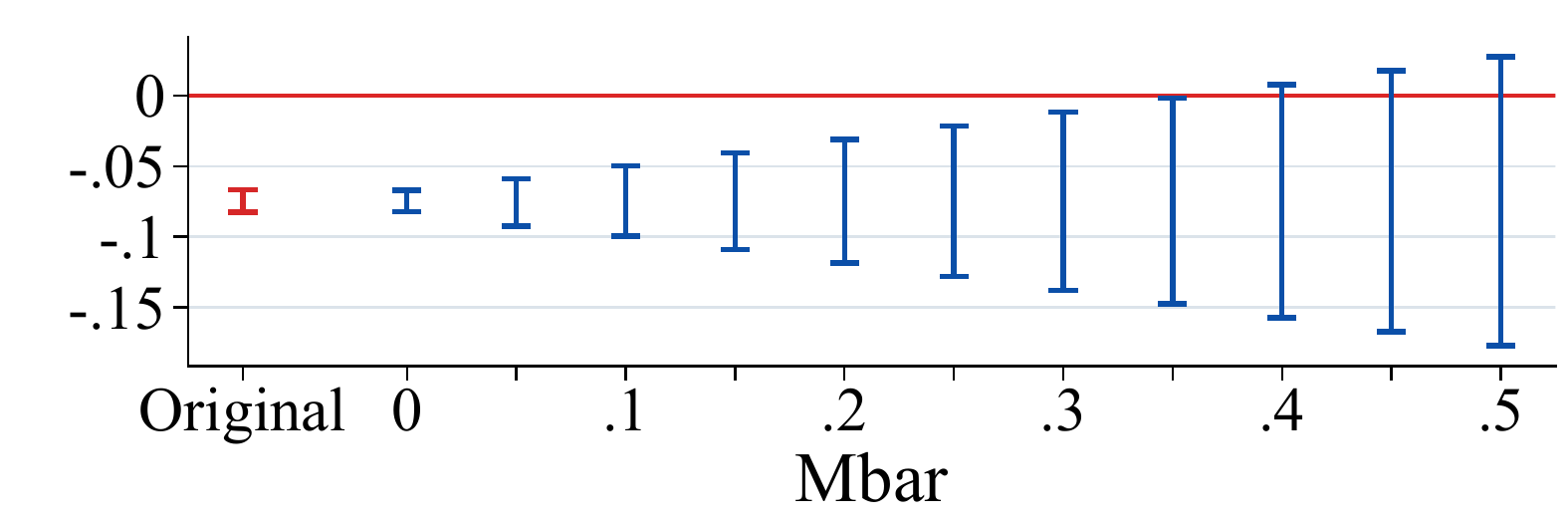} \\
\end{tabular}
\endgroup
\caption{Honest DiD sensitivity intervals under relative-magnitude ($\Delta^{RM}$) restrictions for German and French controls. The target is the equally weighted average of May--December 2024, using the original PPML event-study covariance matrices exported by Stata.}
\label{fig:clickstream_honestdid_rm}
\end{figure}

We use the relative-magnitude restriction, $\Delta^{RM}(\bar M)$, as the observed pre-treatment deviations appear primarily as fluctuations around zero rather than as a stable, negative differential trend. A restriction calibrated to the magnitude of adjacent pre-treatment changes is therefore more natural in this setting. Let
$\mathcal{M}_{\mathrm{pre}}^{0}
=
\mathcal{M}_{\mathrm{pre}}\cup\{m_0\}$.
The relative-magnitude restriction requires
\begin{equation}
\label{eq:clickstream_honest_did_rm}
\left|
\delta_{c,r}-\delta_{c,r-1}
\right|
\leq
\bar M
\max_{\substack{
s\in\mathcal{M}_{\mathrm{pre}}^{0}\\
s-1\in\mathcal{M}_{\mathrm{pre}}^{0}
}}
\left|
\delta_{c,s}-\delta_{c,s-1}
\right|,
\qquad
r\in\mathcal{M}_{\mathrm{post}}.
\end{equation}
Thus, $\bar M$ bounds each adjacent post-treatment change in the counterfactual differential trend relative to the largest adjacent pre-treatment change, including the change into the normalized reference month. For instance, $\bar M=0.3$ allows each post-treatment change to be at most $30\%$ as large as the {\it largest} corresponding pre-treatment change. In this way, Honest DiD constructs confidence intervals for the equally weighted May--December 2024 average, accounting for sampling uncertainty in both the pre-treatment differential trends and the post-treatment event-study estimates. Figure~\ref{fig:clickstream_honestdid_rm} reports the relative-magnitude $\Delta^{RM}$ intervals.

For the German comparison, the robust interval remains below zero through $\bar M=0.40$, and for the French comparison, it remains below zero through $\bar M=0.35$. Considering eight post-treatment periods, each with the possibility of deviating by $0.4$ or $0.35$ or the largest pre-treatment deviation, these $\bar{M}$ values represent considerable robustness in our results.

\newcommand{\clickstreamlinearsectiontitle}{Linear Extrapolation of Pre-Treatment Event-Study Coefficients}

\subsection{\clickstreamlinearsectiontitle}
\label{appssec:clickstream_event_study_linear_counterfactual}
As a complementary diagnostic, we utilize a linear extrapolation method in this section. In our linear extrapolation method, we fit a straight line to the pre-treatment event-study coefficients using OLS and WLS methods, extrapolate it to the post-treatment period as the counterfactual trend, and compare the resulting post-treatment mean with the post-treatment event-study coefficients' mean. Let $\hat\beta_{c,r}$ denote the estimated English-control event-study coefficient for control language $c$ in event month $r$. Using only the pre-treatment event months, we fit
\begin{equation}
\label{eq:clickstream_event_study_linear_fit}
\hat\beta_{c,r}
=
\omega_c(r-m_0)+u_{c,r},
\qquad
r\in\mathcal{M}_{\mathrm{pre}}\setminus\{m_0\},
\end{equation}
where $\mathcal{M}_{\mathrm{pre}}$ is the set of pre-treatment event months, $m_0=-1$ is the omitted April 2024 reference month, $\omega_c$ is the pre-treatment slope in PPML log points per event month for control $c$, and $u_{c,r}$ is the corresponding fitting residual. Because the event-study normalization implies $\hat\beta_{c,m_0}=0$, Equation~\eqref{eq:clickstream_event_study_linear_fit} has no free intercept: the line is anchored at $(m_0,0)$. We estimate the equation separately for the German and French controls.

The fitted counterfactual path in a post-treatment event month $r\in\mathcal{M}_{\mathrm{post}}$ is $\hat\beta_{c,r}^{\mathrm{cf}}=\hat\omega_c(r-m_0)$. We therefore report three post-treatment quantities
\begin{equation}
\label{eq:clickstream_event_study_post_summaries}
\widehat{\bar\beta}_c
=
\frac{1}{|\mathcal{M}_{\mathrm{post}}|}
\sum_{r\in\mathcal{M}_{\mathrm{post}}}\hat\beta_{c,r},
\qquad
\widehat{\bar\beta}_c^{\mathrm{cf}}
=
\frac{1}{|\mathcal{M}_{\mathrm{post}}|}
\sum_{r\in\mathcal{M}_{\mathrm{post}}}\hat\omega_c(r-m_0),
\qquad
\widehat{\Delta}_c
=
\widehat{\bar\beta}_c-\widehat{\bar\beta}_c^{\mathrm{cf}}.
\end{equation}
Here $\widehat{\bar\beta}_c$ is the factual post-treatment mean, $\widehat{\bar\beta}_c^{\mathrm{cf}}$ is the counterfactual mean by extrapolating the fitted pre-treatment line, and $\widehat{\Delta}_c$ is their difference. We report two estimators of Equation~\eqref{eq:clickstream_event_study_linear_fit}: ordinary least squares (OLS), which weights the pre-treatment coefficients equally, and weighted least squares (WLS), which uses inverse marginal-variance weights. For each estimator, Table~\ref{tab:clickstream_event_study_linear_counterfactual} reports $\hat\omega_c$, $\widehat{\bar\beta}_c$, $\widehat{\bar\beta}_c^{\mathrm{cf}}$, and $\widehat{\Delta}_c$.

For every control and both fitting methods, the factual post-treatment mean is significantly below the extrapolated counterfactual mean. The estimated differences are German OLS: -0.0510; German WLS: -0.0505; French OLS: -0.1163; French WLS: -0.1100. This diagnostic is consistent with a negative post-AIO shift under a linear continuation of the pre-treatment differential.

\begin{figure}[htp!]
\centering
\paperassetgraphic[width=0.98\textwidth,keepaspectratio]{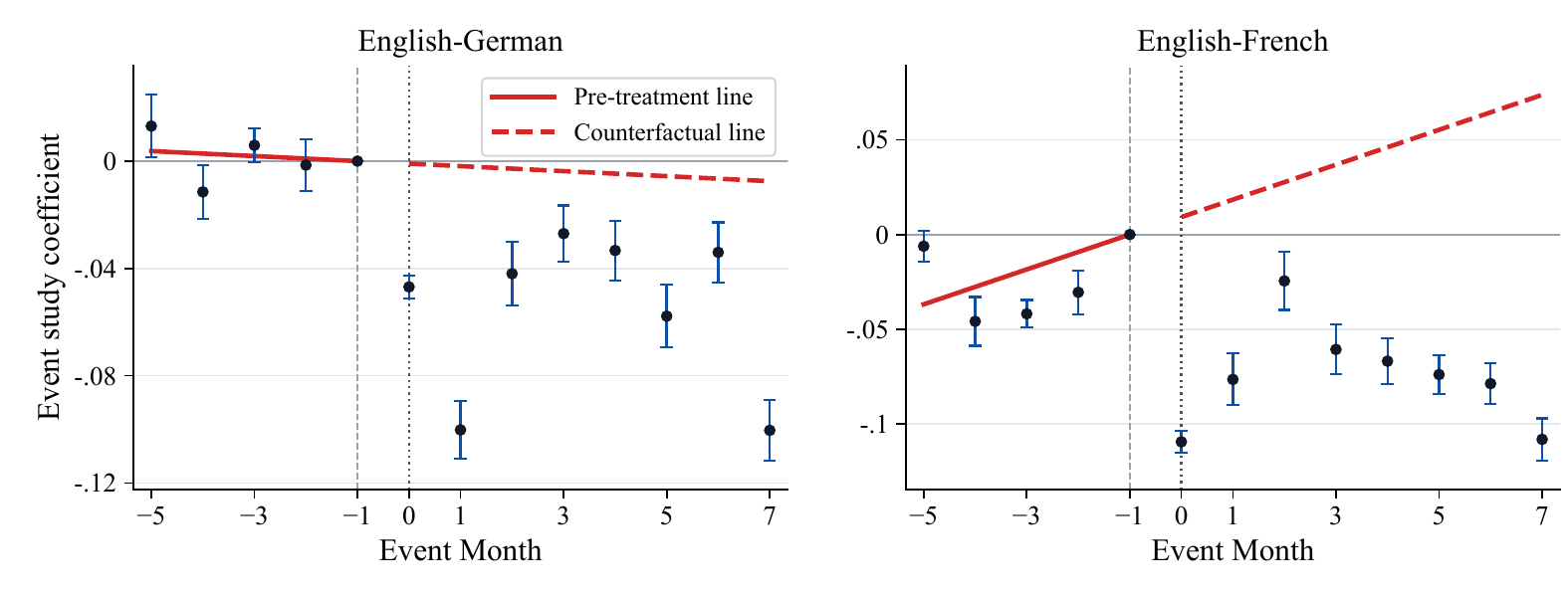}
\caption{OLS linear extrapolations of the pre-treatment event-study coefficients. Black points and blue whiskers report the monthly PPML estimates and 95\% confidence intervals. The solid red line is fitted to the four nonreference pre-treatment coefficients and anchored at the omitted April 2024 coefficient; the dashed red line is its May--December counterfactual continuation. The dotted gray line marks May 2024.}
\label{fig:clickstream_linear_counterfactual_ols}
\end{figure}

\begin{figure}[htp!]
\centering
\paperassetgraphic[width=0.98\textwidth,keepaspectratio]{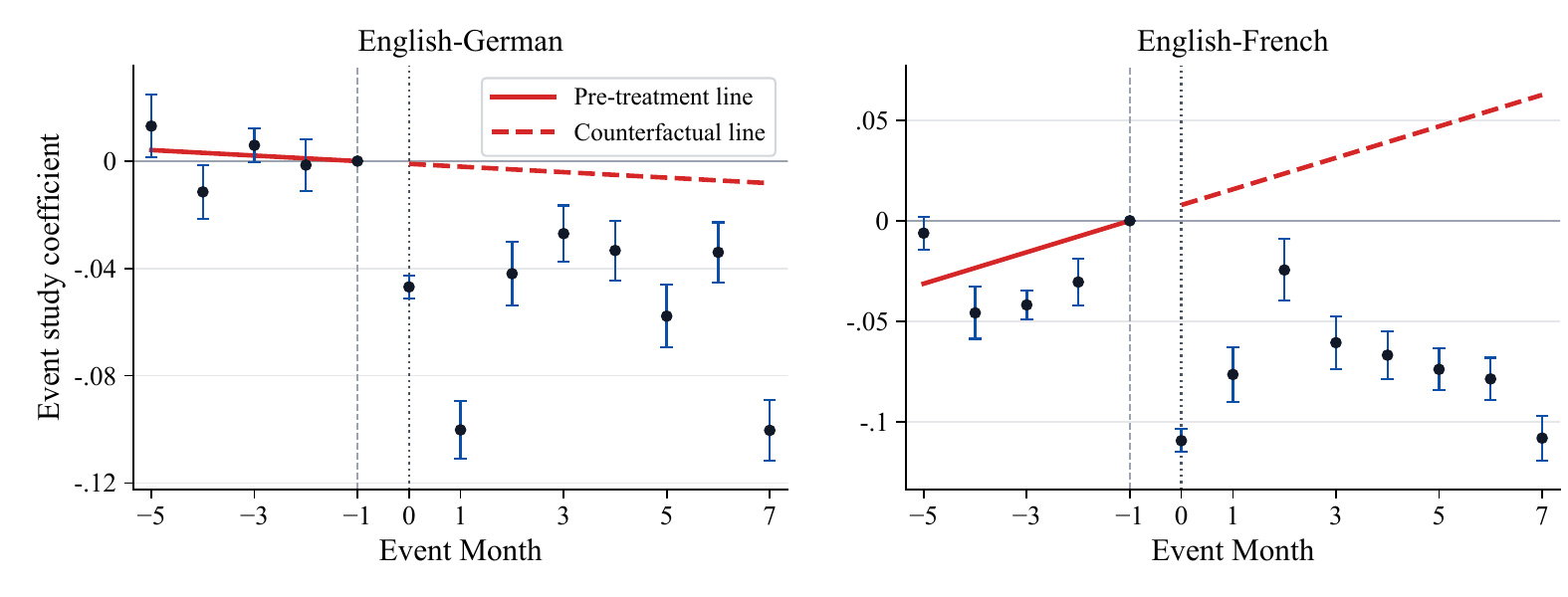}
\caption{WLS linear extrapolations of the pre-treatment event-study coefficients. Black points and blue whiskers report the monthly PPML estimates and 95\% confidence intervals. The solid red line uses inverse marginal-variance weights for the four nonreference pre-treatment coefficients and is anchored at the omitted April 2024 coefficient; the dashed red line is its May--December counterfactual continuation. The dotted gray line marks May 2024.}
\label{fig:clickstream_linear_counterfactual_wls}
\end{figure}

Figures~\ref{fig:clickstream_linear_counterfactual_ols} and~\ref{fig:clickstream_linear_counterfactual_wls}   plot the monthly event-study estimates together with the fitted OLS and WLS pre-treatment paths and their post-treatment counterfactual continuations.

\FloatBarrier

\begin{table}[htp!]
\centering
\caption{Linear extrapolations of pre-treatment event-study coefficients}
\label{tab:clickstream_event_study_linear_counterfactual}
\small
\setlength{\tabcolsep}{5pt}
\begin{tabular}{llcccc}
\toprule
Control & Method & Pre-trend slope & Factual post mean & Counterfactual post mean & Difference \\
\midrule
German & OLS & $-0.0009$ & $-0.0552^{***}$ & $-0.0042$ & $-0.0510^{***}$ \\
 & & (0.0013) & (0.0039) & (0.0057) & (0.0079) \\ 
 & WLS & $-0.0010$ & $-0.0552^{***}$ & $-0.0046$ & $-0.0505^{***}$ \\
 & & (0.0013) & (0.0039) & (0.0057) & (0.0078) \\ 
\midrule
French & OLS & $0.0092^{***}$ & $-0.0748^{***}$ & $0.0414^{***}$ & $-0.1163^{***}$ \\
 & & (0.0012) & (0.0040) & (0.0052) & (0.0072) \\ 
 & WLS & $0.0078^{***}$ & $-0.0748^{***}$ & $0.0352^{***}$ & $-0.1100^{***}$ \\
 & & (0.0011) & (0.0040) & (0.0049) & (0.0070) \\ 
\bottomrule
\end{tabular}
\begin{minipage}{0.98\linewidth}
\vspace{0.35em}\small
\textit{Notes:} Estimates are PPML log coefficients. OLS gives equal weight to the four nonreference pre-treatment coefficients, and WLS uses inverse marginal-variance weights. The fitted line is anchored at the omitted April 2024 reference month. ``Difference'' is the factual May--December 2024 mean minus the extrapolated counterfactual mean. Standard errors computed using the unaltered event-study covariance matrix are in parentheses. $^{***}p<0.001$, $^{**}p<0.01$, $^{*}p<0.05$.
\end{minipage}
\end{table}

\clearpage

\section{Sensitivity to Post-Treatment Censoring Assignment}
\label{appsec:clickstream_imputation_sensitivity}

The main analysis addresses unpublished post-treatment counts by setting an unpublished English count to 10, and an unpublished control count to 0. This assignment moves censored outcomes against a negative English effect. To assess how strongly the estimates depend on that choice, we run the PPML regression in Equation~\eqref{eq:pooled_ppml_did} on three alternative imputation procedures: we first reverse the assignment for censored post-treatment observations, setting an unpublished English count to 0 and an unpublished control count to 10. Alternatively, we assign censored observations a value of 10 in both the English and control editions, or assign a value of 0 in both editions. Table~\ref{tab:clickstream_imputation_sensitivity} reports the main and alternative assignment estimates.

\begin{table}[htp!]
\centering
\small
\setlength{\tabcolsep}{3.0pt}
\begin{threeparttable}
\caption{PPML DiD estimates under alternative post-treatment censoring assignments}
\label{tab:clickstream_imputation_sensitivity}
\begin{tabular}{lcc}
\toprule
 & German & French \\
\midrule
\multicolumn{3}{l}{\textit{Main assignment: unpublished English count 10; control count 0}} \\
English $\times$ post & $-0.0561^{***}$ & $-0.0494^{***}$ \\
 & (0.0117) & (0.0136) \\
\addlinespace[0.45em]
\multicolumn{3}{l}{\textit{Opposite assignment: unpublished English count 0; control count 10}} \\
English $\times$ post & $-0.0566^{***}$ & $-0.0502^{***}$ \\
 & (0.0117) & (0.0136) \\
\addlinespace[0.45em]
\multicolumn{3}{l}{\textit{Symmetric zero assignment: unpublished counts 0 in both languages}} \\
English $\times$ post & $-0.0561^{***}$ & $-0.0494^{***}$ \\
 & (0.0117) & (0.0136) \\
\addlinespace[0.45em]
\multicolumn{3}{l}{\textit{Symmetric ten assignment: unpublished counts 10 in both languages}} \\
English $\times$ post & $-0.0566^{***}$ & $-0.0502^{***}$ \\
 & (0.0117) & (0.0136) \\
\addlinespace[0.45em]
No. observations & 12,998,102 & 13,802,698 \\
Articles (QIDs) & 499,927 & 530,873 \\
\midrule
Article $\times$ language FE & Yes & Yes \\
Month FE & Yes & Yes \\
\bottomrule
\end{tabular}
\begin{tablenotes}[flushleft]\small
\item Published counts are unchanged under both assignments. Robust standard errors in parentheses are two-way clustered by article and month.
\item Significance: $^{***}p<0.001$, $^{**}p<0.01$, $^{*}p<0.05$.
\end{tablenotes}
\end{threeparttable}
\end{table}

\FloatBarrier
Under every assignment, the estimates imply a near 5.5\% decline relative to German and a near 4.9\% decline relative to French. All estimates are statistically significant. The negative direction therefore does not depend on which of the two directional assignments we use.

\clearpage

\section{Pre-Treatment Timing Placebo}
\label{appsec:clickstream_preperiod_timing_placebo}

In this section, we run a placebo analysis to check whether moving the treatment date to two months earlier and using only the pre-treatment period (December 2023--April 2024) yields similar negative estimates. We define $\text{PlaceboPost}_m$ to equal one in March and April 2024 (placebo post-treatment period) and zero from December 2023 through February 2024 (placebo pre-treatment period). We then estimate
\begin{equation}
\label{eq:clickstream_preperiod_timing_placebo}
\mathbb{E}\!\left[R_{i,l,m}\mid \alpha_{i,l},\delta_m\right]
=
\exp\!\left(
\alpha_{i,l}
+
\delta_m
+
\beta_c^{\mathrm{P}}
\big(\text{English}_l\times\text{PlaceboPost}_m\big)
\right),
\end{equation}
separately for each $c\in\{\mathrm{DE},\mathrm{FR}\}$. The article--language and month fixed effects are identical to those in Equation~\eqref{eq:pooled_ppml_did}. The coefficient $\beta_c^{\mathrm{P}}$ measures whether English search traffic changed relative to control $c$ after the placebo March onset.

If the main estimates capture a post-AIO shift rather than an existing English--control divergence, $\beta_c^{\mathrm{P}}$ should be close to zero. A negative placebo estimate would instead indicate that the relative decline had already begun before May. Because sample membership requires published counts in all five months, this exercise tests timing within the retained sample. Table~\ref{tab:clickstream_preperiod_timing_placebo} reports the placebo estimates.

\IfFileExists{tables/clickstream/ppml_did_preperiod_timing_placebo.tex}{%
\begin{table}[htp!]
\centering
\small
\setlength{\tabcolsep}{2.0pt}
\begin{threeparttable}
\caption{Pre-treatment timing-placebo PPML DiD estimates}
\label{tab:clickstream_preperiod_timing_placebo}
\begin{tabular}{lcc}
\toprule
 & German & French \\
\midrule
English $\times$ placebo post & $-0.0026$ & $0.0162$ \\
 & (0.0094) & (0.0179) \\[0.25em]
No. observations & 4,999,270 & 5,308,730 \\
Articles (QIDs) & 499,927 & 530,873 \\
\midrule
Article $\times$ language FE & Yes & Yes \\
Month FE & Yes & Yes \\
\bottomrule
\end{tabular}
\begin{tablenotes}[flushleft]\small
\item The placebo pre-period is December 2023--February 2024, and the placebo post-period is March--April 2024. Robust standard errors in parentheses are two-way clustered by article and month.
\item Significance: $^{***}p<0.001$, $^{**}p<0.01$, $^{*}p<0.05$.
\end{tablenotes}
\end{threeparttable}
\end{table}
}{}
\FloatBarrier

\IfFileExists{tables/clickstream/ppml_did_preperiod_timing_placebo_interpretation.tex}{%
Neither placebo estimate is statistically significant. Thus, the timing falsification provides no evidence of the negative English-relative-control shift before the actual treatment period.
}{}

\clearpage

\section{English--Japanese Analysis Through July 2024}
\label{appsec:japanese_control}

In this section, we report the English--Japanese comparison separately from the two main control comparisons. Japan entered Google's public AIO expansion in August 2024 \citep{Google2024AIOAugExpansion}; ending the outcome window in July therefore keeps this comparison entirely before that expansion. The sample-membership and post-treatment censoring rules are otherwise identical to the main analysis: both editions must have published counts in every pre-treatment month, and post-treatment publication does not affect eligibility. All descriptive outcomes, model-free series, and estimates reported below use December 2023--July 2024. The resulting sample contains \clickMainJAPairs\ matched English--Japanese article pairs.

\subsection{Summary Statistics and Model-Free Series}

Table~\ref{tab:sumstats_japanese} reports the pre-treatment (December 2023--April 2024) and post-treatment (May--July 2024) distributions of monthly article-level search traffic in the English--Japanese sample.

\begin{table}[htp!]
\centering
\small
\caption{Summary statistics of monthly search traffic for the English--Japanese sample through July 2024.}
\label{tab:sumstats_japanese}
\setlength{\tabcolsep}{2pt}%
\makebox[\textwidth][c]{%
\begin{tabular}{lcc}
\toprule
\midrule
\multicolumn{3}{c}{Pre-treatment months (Dec.~2023 -- Apr.~2024)} \\
\midrule
Statistic & English (EN--JA) & Japanese \\
\midrule
Mean & 5,836.47 & 926.30 \\
Std. & 31,743.35 & 5,530.86 \\
Min. & 10.00 & 10.00 \\
25\% & 253.00 & 45.00 \\
50\% & 1,034.00 & 133.00 \\
75\% & 3,865.00 & 503.00 \\
Max. & 12,503,258.00 & 1,580,324.00 \\
Article-Language & 315,054 & 315,054 \\
\midrule
\multicolumn{3}{c}{Post-treatment months (May~2024 -- Jul.~2024)} \\
\midrule
Statistic & English (EN--JA) & Japanese \\
\midrule
Mean & 4,946.87 & 940.54 \\
Std. & 24,905.92 & 4,835.75 \\
Min. & 10.00 & 0.00 \\
25\% & 215.00 & 48.00 \\
50\% & 885.00 & 145.00 \\
75\% & 3,310.00 & 547.00 \\
Max. & 8,113,164.00 & 1,039,675.00 \\
Article-Language & 315,054 & 315,054 \\
\bottomrule
\end{tabular}
}
\end{table}

\begin{figure}[htp!]
\centering
\small
English--Japanese\\[0.25em]
\paperassetgraphic[width=0.58\textwidth,keepaspectratio]{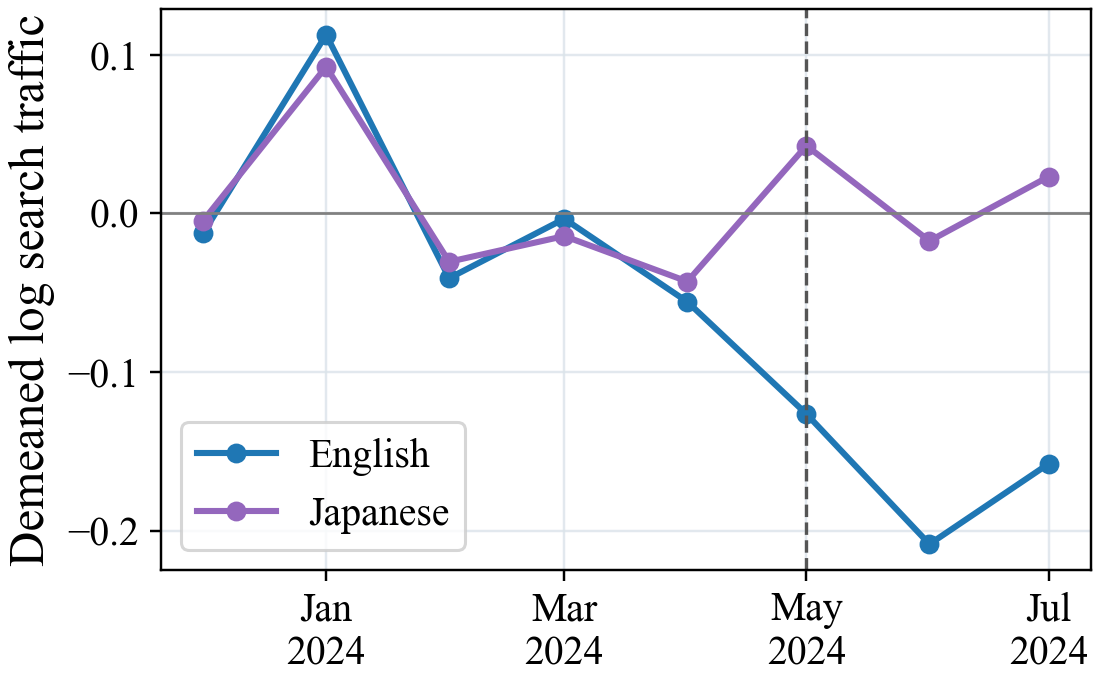}
\caption{Demeaned log total monthly search referrals through July 2024 for the English--Japanese sample. Each series is demeaned using December 2023--April 2024; the dashed line marks May 2024.}
\label{fig:clickstream_model_free_japanese}
\end{figure}

Figure~\ref{fig:clickstream_model_free_japanese} applies the same model-free transformation as Figure~\ref{fig:clickstream_model_free}: total search traffic in each language is transformed using $\log(1+R)$ and demeaned by the language-specific December 2023--April 2024 mean.

\FloatBarrier

\subsection{PPML Difference-in-Differences Estimate}

We estimate Equation~\eqref{eq:pooled_ppml_did} on the English--Japanese sample through July 2024. The coefficient of $\clickMainJABeta$ implies that monthly English Wikipedia search traffic declined by approximately $\clickMainJADecline$ relative to Japanese beginning in May 2024. Table~\ref{tab:clickstream_ppml_did_japanese} reports the coefficient and two-way-clustered standard error.

\begin{table}[htp!]
\centering
\small
\setlength{\tabcolsep}{2.0pt}
\begin{threeparttable}
\caption{English--Japanese PPML DiD estimate through July 2024}
\label{tab:clickstream_ppml_did_japanese}
\begin{tabular}{lc}
\toprule
 & Japanese \\
\midrule
English $\times$ post & $-0.1806^{***}$ \\
 & (0.0110) \\[0.25em]
No. observations & 5,040,864 \\
Articles (QIDs) & 315,054 \\
\midrule
Article $\times$ language FE & Yes \\
Month FE & Yes \\
\bottomrule
\end{tabular}
\begin{tablenotes}[flushleft]\small
\item Robust standard errors in parentheses are two-way clustered by article and month.
\item Significance: $^{***}p<0.001$, $^{**}p<0.01$, $^{*}p<0.05$.
\end{tablenotes}
\end{threeparttable}
\end{table}

\subsection{Parallel-Trends and Sensitivity Diagnostics}

In this section, we apply the same sequence of diagnostics as in Web Appendix~$\S$\ref{appsec:clickstream_parallel_pretrends} to assess the parallel trends assumption on the English-Japanese sample. 

\noindent \paragraph{Linear pre-treatment test.} Table~\ref{tab:clickstream_ppml_pretrend_japanese} reports the PPML differential linear trend over the five pre-treatment months. The estimated coefficient is negative but statistically insignificant, representing no differential trend over time between the search traffic to English and Japanese articles.

\begin{table}[htp!]
\centering
\small
\setlength{\tabcolsep}{2.0pt}
\begin{threeparttable}
\caption{English--Japanese PPML linear pre-treatment trend test}
\label{tab:clickstream_ppml_pretrend_japanese}
\begin{tabular}{lc}
\toprule
 & Japanese \\
\midrule
English $\times t_m$ & $-0.0020$ \\
 & (0.0049) \\[0.25em]
No. observations & 3,150,540 \\
Articles (QIDs) & 315,054 \\
\midrule
Article $\times$ language FE & Yes \\
Month FE & Yes \\
\bottomrule
\end{tabular}
\begin{tablenotes}[flushleft]\small
\item Robust standard errors in parentheses are two-way clustered by article and month.
\item Significance: $^{***}p<0.001$, $^{**}p<0.01$, $^{*}p<0.05$.
\end{tablenotes}
\end{threeparttable}
\end{table}

\begin{figure}[htp!]
\centering
\small
English--Japanese\\[0.25em]
\paperassetgraphic[width=0.58\textwidth,keepaspectratio]{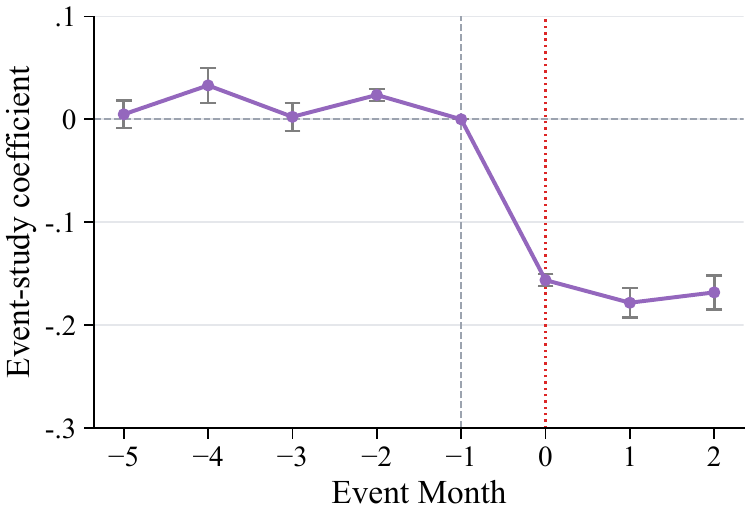}
\caption{Monthly PPML event-study estimates through July 2024 for the English--Japanese sample. April 2024 is the omitted reference month and May 2024 is event month zero. Pointwise 95\% confidence intervals use the direct two-way-clustered standard errors and asymptotic-normal inference from Stata.}
\label{fig:clickstream_event_study_japanese}
\end{figure}

\noindent \paragraph{Event-study analysis.} Figure~\ref{fig:clickstream_event_study_japanese} reports the monthly event-study coefficients through July 2024, with April 2024 omitted and May 2024 defined as event month zero. The estimated coefficients are close to zero, and in cases where the 95\% confidence intervals do not include zero, the deviations are much smaller compared to the post-treatment effect. Next, we present the relative-magnitude Honest DiD results to assess the sensitivity of the English-Japanese PPML difference-in-differences result to deviations from pre-treatment parallel trends.

\begin{figure}[htp!]
\centering
\small
English--Japanese\\[0.25em]
\paperassetgraphic[width=0.58\textwidth,keepaspectratio]{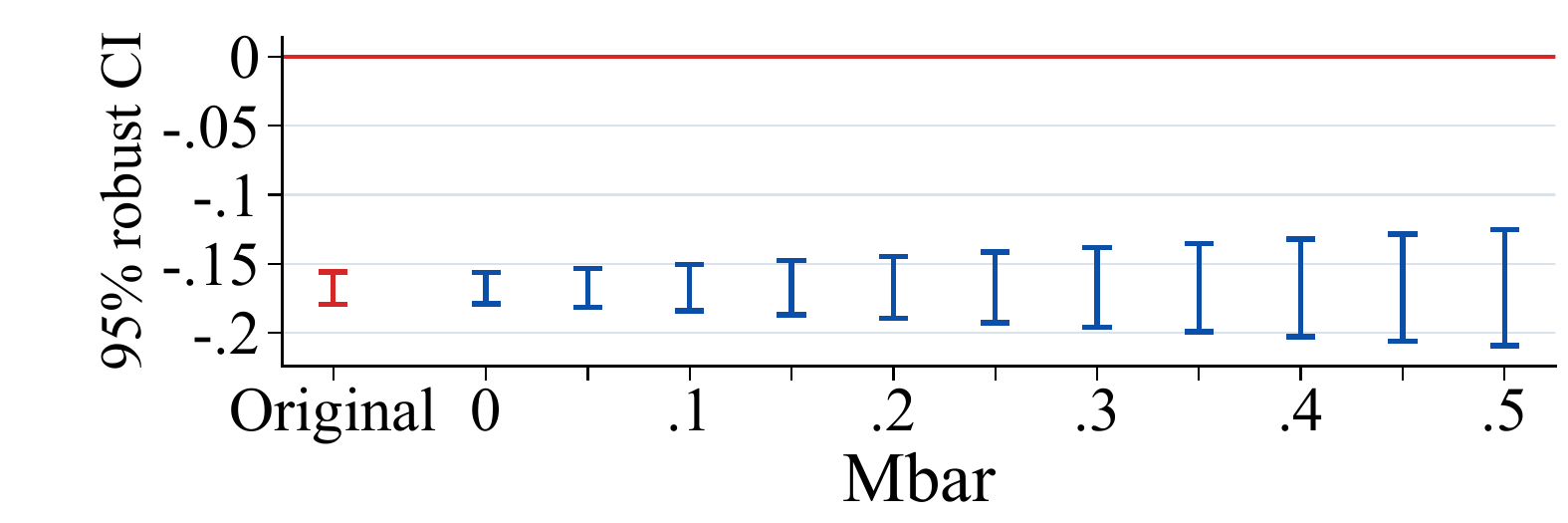}
\caption{Honest DiD sensitivity intervals under relative-magnitude ($\Delta^{RM}$) restrictions for the Japanese control. The target is the equally weighted average of May--July 2024.}
\label{fig:clickstream_honestdid_rm_japanese}
\end{figure}

\noindent \paragraph{Honest DiD.} Figure~\ref{fig:clickstream_honestdid_rm_japanese} reports Honest DiD intervals under the relative-magnitude restriction in Equation~\eqref{eq:clickstream_honest_did_rm}. The target is the equally weighted May--July 2024 average. All the 95\% confidence intervals are below zero, showing the relative robustness of the negative result in Table~\ref{tab:clickstream_ppml_did_japanese}

\begin{figure}[htp!]
\centering
\begin{subfigure}[t]{0.49\textwidth}
\centering
\paperassetgraphic[width=\linewidth,keepaspectratio]{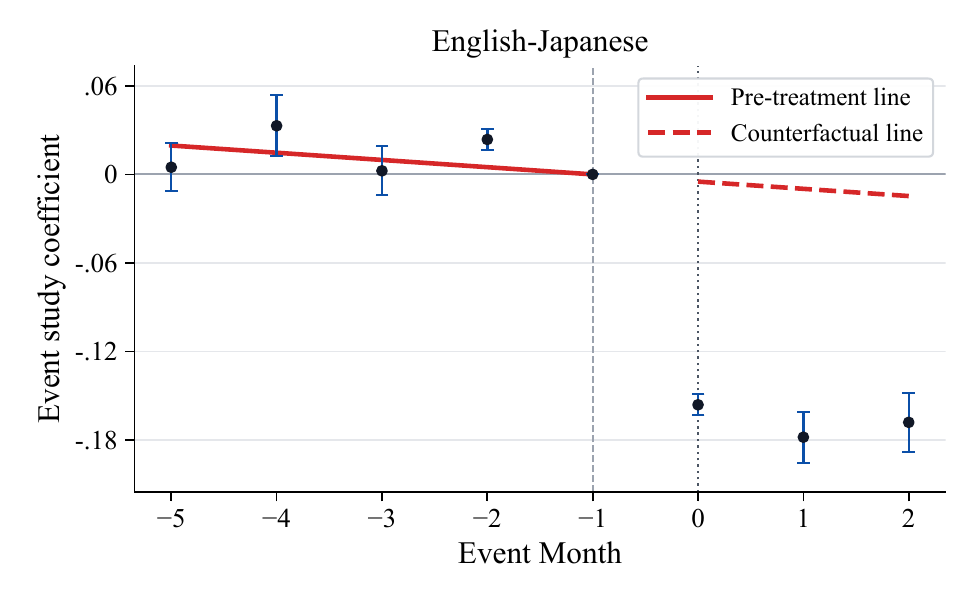}
\caption{OLS}
\label{fig:clickstream_linear_counterfactual_japanese_ols}
\end{subfigure}\hfill
\begin{subfigure}[t]{0.49\textwidth}
\centering
\paperassetgraphic[width=\linewidth,keepaspectratio]{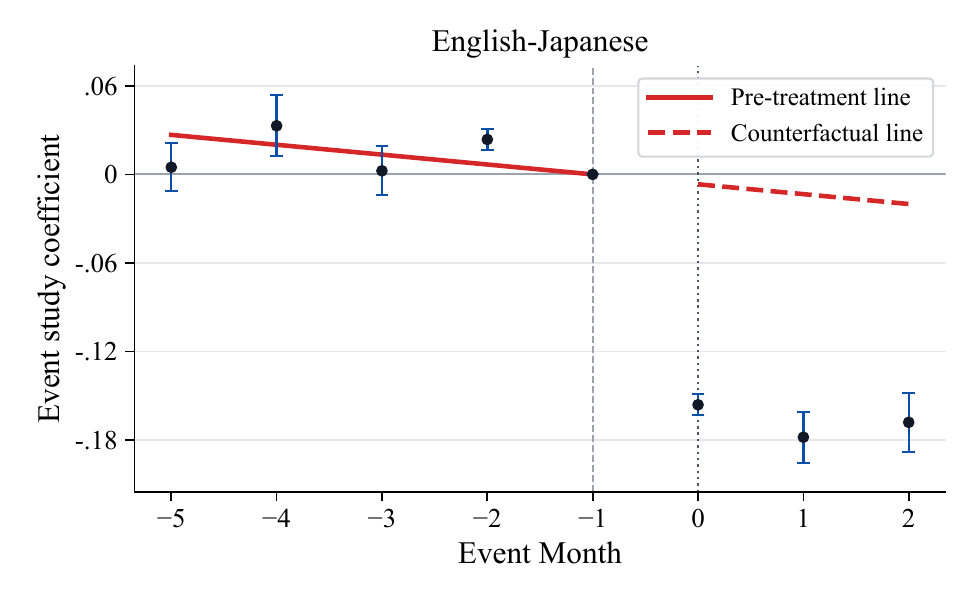}
\caption{WLS}
\label{fig:clickstream_linear_counterfactual_japanese_wls}
\end{subfigure}
\caption{Linear extrapolations for the English--Japanese comparison through July 2024. Black points and blue whiskers report the monthly PPML estimates and 95\% confidence intervals. The solid red line is fitted to the four nonreference pre-treatment coefficients and anchored at the omitted April 2024 coefficient, using equal weights in panel (a) and inverse marginal-variance weights in panel (b); the dashed red line is its May--July counterfactual continuation. The dotted gray line marks May 2024.}
\end{figure}

\noindent \paragraph{Linear Counterfactual Extrapolation}Figures~\ref{fig:clickstream_linear_counterfactual_japanese_ols} and~\ref{fig:clickstream_linear_counterfactual_japanese_wls} report the corresponding OLS and WLS pre-treatment paths and their May--July counterfactual continuations for the English--Japanese comparison.

\FloatBarrier

Finally, Table~\ref{tab:clickstream_event_study_linear_counterfactual_japanese} fits the pre-treatment event-study coefficients using the anchored OLS and WLS specifications in Equations~\eqref{eq:clickstream_event_study_linear_fit}--\eqref{eq:clickstream_event_study_post_summaries} and compares the factual May--July average with the extrapolated counterfactual average.

\begin{table}[htp!]
\centering
\caption{Linear extrapolations of pre-treatment event-study coefficients}
\label{tab:clickstream_event_study_linear_counterfactual_japanese}
\small
\setlength{\tabcolsep}{5pt}
\begin{tabular}{llcccc}
\toprule
Control & Method & Pre-trend slope & Factual post mean & Counterfactual post mean & Difference \\
\midrule
Japanese & OLS & $-0.0049^{*}$ & $-0.1675^{***}$ & $-0.0098^{*}$ & $-0.1577^{***}$ \\
 & & (0.0022) & (0.0059) & (0.0044) & (0.0085) \\ 
 & WLS & $-0.0067^{***}$ & $-0.1675^{***}$ & $-0.0134^{***}$ & $-0.1541^{***}$ \\
 & & (0.0018) & (0.0059) & (0.0036) & (0.0073) \\ 
\bottomrule
\end{tabular}
\begin{minipage}{0.98\linewidth}
\vspace{0.35em}\small
\textit{Notes:} OLS gives equal weight to the four nonreference pre-treatment coefficients, and WLS uses inverse marginal-variance weights. The fitted line is anchored at the omitted April 2024 reference month. ``Difference'' is the factual May--July 2024 mean minus the extrapolated counterfactual mean. Standard errors computed using the unaltered event-study covariance matrix are in parentheses. $^{***}p<0.001$, $^{**}p<0.01$, $^{*}p<0.05$.
\end{minipage}
\end{table}

For both fitting methods, the factual post-treatment mean is significantly below the extrapolated counterfactual mean. The estimated differences are OLS: -0.1577; WLS: -0.1541. This diagnostic is consistent with a negative post-AIO shift under a linear continuation of the pre-treatment differential.

\clearpage

\section{Traffic and Advertising-Revenue Calculations}
\label{appsec:economic_magnitude}

This section provides the calculations underlying the traffic and
advertising-revenue benchmarks discussed in
Section~\ref{sec:discussion_implications}. We use these calculations to
illustrate the economic scale of the main estimates. Because we estimate the
English--German and English--French specifications on separately constructed
article samples, the calculations do not constitute a separate pooled causal
estimate.

\subsection{Implied Reduction in Direct Search Traffic}

Table~\ref{tab:clickstream_ppml_did} reports the English--German and
English--French PPML coefficients,
$\hat{\beta}_{\mathrm{DE}}=\clickMainDEBeta$ and
$\hat{\beta}_{\mathrm{FR}}=\clickMainFRBeta$. To construct a single
back-of-the-envelope benchmark, we take their arithmetic mean:

\begin{equation}
\label{eq:appendix_average_ppml_coefficient}
\hat{\beta}_{\mathrm{avg}}
=
\frac{1}{2}
\left(
\hat{\beta}_{\mathrm{DE}}
+
\hat{\beta}_{\mathrm{FR}}
\right)
=
\clickMainEconomicBeta.
\end{equation}

This coefficient corresponds to an average proportional decline of
approximately \clickMainEconomicDecline. We match this average coefficient with a corresponding average traffic base.
Let
$\overline{T}^{\mathrm{EN,DE}}_{\mathrm{post}}$ and
$\overline{T}^{\mathrm{EN,FR}}_{\mathrm{post}}$ denote average monthly English
search traffic from May through December 2024 in the English--German and
English--French samples, respectively. We define

\begin{equation}
\label{eq:appendix_average_post_english_traffic}
\overline{T}^{\mathrm{EN}}_{\mathrm{post,avg}}
=
\frac{1}{2}
\left(
\overline{T}^{\mathrm{EN,DE}}_{\mathrm{post}}
+
\overline{T}^{\mathrm{EN,FR}}_{\mathrm{post}}
\right)
\approx
\clickMainFRPostAverageBillions
\ \text{billion visits per month}.
\end{equation}

The observed post-treatment traffic already incorporates the estimated
decline. Under the multiplicative PPML model, the corresponding
counterfactual traffic level absent the estimated effect equals
$\overline{T}^{\mathrm{EN}}_{\mathrm{post,avg}}
\exp(-\hat{\beta}_{\mathrm{avg}})$. We therefore calculate the implied monthly
traffic loss as

\begin{equation}
\label{eq:appendix_monthly_search_traffic_loss}
\widehat{L}_{\mathrm{month}}
=
\overline{T}^{\mathrm{EN}}_{\mathrm{post,avg}}
\left[
\exp\!\left(-\hat{\beta}_{\mathrm{avg}}\right)-1
\right]
\approx
\clickMainMonthlyLossMillions
\ \text{million visits per month}.
\end{equation}

Thus, the average of the two main estimates implies approximately
\clickMainMonthlyLossMillions\ million fewer direct search-originated visits
to English Wikipedia each month. If the estimated monthly rate persisted for
one year, the corresponding annualized reduction would be

\begin{equation}
\label{eq:appendix_annual_search_traffic_loss}
\widehat{L}_{\mathrm{year}}
=
12\widehat{L}_{\mathrm{month}}
\approx
\clickMainAnnualLossBillions
\ \text{billion visits per year}.
\end{equation}

The annualized value is an illustrative scenario rather than an estimate over
an observed twelve-month treatment period. It assumes that the average
post-treatment effect persists despite subsequent changes in AIO exposure,
user behavior, and publisher responses. Moreover, the calculation includes
only visits referred directly by external search engines; it does not include
any associated changes in subsequent internal navigation within Wikipedia.

\subsection{Illustrative Advertising-Revenue Equivalent}

Wikipedia does not display advertising, so the estimated traffic reduction
does not imply an advertising-revenue loss for Wikipedia itself. To illustrate
the potential scale for an otherwise comparable ad-supported publisher, we
assume that each foregone search referral corresponds to one foregone
landing-page view and apply benchmark display-ad revenues per page.

\citet{ZengKohnoRoesner2020BadNews} report an average of 5.22 display
advertisements on mainstream news article pages, which we use as an
illustrative measure of ad opportunities per page. We assume a fill rate
between 75\% and 90\%, where the fill rate denotes the share of ad requests
that result in an impression \citep{GoogleAdManagerFillRate}. The upper portion
of this range is consistent with benchmarks for publishers with relatively
strong advertising monetization
\citep{Playwire2026StatePublisherAdRevenue}. We further assume that publishers
retain 70\% of advertiser spending, close to the 73.9\% seller-revenue share
reported in the ANA programmatic-advertising benchmark
\citep{ANA2025ProgrammaticBenchmarkQ1}.

The ANA benchmark reports median CPMs of \$3.28 in the open marketplace and
\$9.37 in private marketplaces
\citep{ANA2025ProgrammaticBenchmarkQ1}. Because the benchmark covers multiple
delivery environments, including connected television, we treat these values
as illustrative low and high inputs rather than as general bounds for
web-display CPMs.

Combining the lower CPM with a 75\% fill rate and the higher CPM with a 90\%
fill rate gives the following publisher revenue per thousand page views:

\begin{align}
\label{eq:appendix_low_page_rpm}
\text{Low page RPM}
&=
5.22
\times
0.75
\times
0.70
\times
\$3.28
\approx
\$8.99,
\\
\label{eq:appendix_high_page_rpm}
\text{High page RPM}
&=
5.22
\times
0.90
\times
0.70
\times
\$9.37
\approx
\$30.81.
\end{align}

For a page-RPM benchmark $r$, we calculate the monthly advertising-revenue
equivalent as

\begin{equation}
\label{eq:appendix_monthly_revenue_general}
\Delta\widehat{\mathrm{Rev}}_{\mathrm{month}}(r)
=
\frac{\widehat{L}_{\mathrm{month}}}{1{,}000}
\times r.
\end{equation}

Applying the low and high page-RPM benchmarks to the estimated monthly traffic
loss yields

\begin{equation}
\label{eq:appendix_monthly_revenue_equivalent}
\Delta\widehat{\mathrm{Rev}}_{\mathrm{month}}
\approx
\$\clickMainMonthlyRevenueLowMillions\text{M}
\ \text{to}\
\$\clickMainMonthlyRevenueHighMillions\text{M per month}.
\end{equation}

If these monthly values persisted for one year, the corresponding annualized
revenue equivalent would be

\begin{equation}
\label{eq:appendix_annual_revenue_equivalent}
\Delta\widehat{\mathrm{Rev}}_{\mathrm{year}}
\approx
\$\clickMainAnnualRevenueLowMillions\text{M}
\ \text{to}\
\$\clickMainAnnualRevenueHighMillions\text{M per year}.
\end{equation}

These calculations require several strong assumptions. In particular, we
treat each foregone search referral as one foregone page view, apply
advertising benchmarks from other publishers to Wikipedia-scale traffic, and
assume that the estimated monthly traffic effect persists when we annualize
the result. We also omit potential effects on subscriptions, lead generation,
commerce, and subsequent internal navigation. The resulting figures should
therefore be interpreted as illustrative advertising-revenue equivalents, not
as estimates of Wikipedia's lost revenue or of revenue transferred from
publishers to Google.

\putbib
\end{bibunit}


\end{appendices}

\begin{thebibliography}{60}
\providecommand{\natexlab}[1]{#1}
\providecommand{\url}[1]{\texttt{#1}}
\expandafter\ifx\csname urlstyle\endcsname\relax
  \providecommand{\doi}[1]{doi: #1}\else
  \providecommand{\doi}{doi: \begingroup \urlstyle{rm}\Url}\fi

\bibitem[Agarwal and Sen(2026)]{agarwal2026impact}
S.~Agarwal and A.~Sen.
\newblock The impact of google ai overviews on publisher traffic and user experience: Evidence from a field experiment.
\newblock 2026.

\bibitem[Angrist and Pischke(2009)]{AngristPischke2009}
J.~D. Angrist and J.-S. Pischke.
\newblock \emph{Mostly Harmless Econometrics: An Empiricist's Companion}.
\newblock Princeton University Press, 2009.

\bibitem[Aral et~al.(2026)Aral, Li, and Zuo]{aral2026rise}
S.~Aral, H.~Li, and R.~Zuo.
\newblock The rise of ai search: Implications for information markets and human judgement at scale.
\newblock \emph{arXiv preprint arXiv:2602.13415}, 2026.

\bibitem[{Association of National Advertisers (ANA)}(2025)]{ANA2025ProgrammaticBenchmarkQ1}
{Association of National Advertisers (ANA)}.
\newblock Programmatic transparency benchmark report (q1 2025).
\newblock \url{https://tinyurl.com/4jc4r2pm}, 2025.

\bibitem[Athey et~al.(2021)Athey, Mobius, and Pal]{athey2021impact}
S.~Athey, M.~Mobius, and J.~Pal.
\newblock The impact of aggregators on internet news consumption.
\newblock Technical report, National Bureau of Economic Research, 2021.

\bibitem[Bertrand et~al.(2004)Bertrand, Duflo, and Mullainathan]{Bertrand2004}
M.~Bertrand, E.~Duflo, and S.~Mullainathan.
\newblock How much should we trust differences-in-differences estimates?
\newblock \emph{The Quarterly Journal of Economics}, 119\penalty0 (1):\penalty0 249--275, 2004.

\bibitem[{BrightEdge}(2024)]{brightedge2024AIO}
{BrightEdge}.
\newblock 10 observations about {AI} overviews after a full month.
\newblock \url{https://www.brightedge.com/blog/10-observations-AI-Overviews-June2024}, 2024.

\bibitem[Budaraju(2024)]{Google2024AIOAugExpansion}
H.~Budaraju.
\newblock New ways to connect to the web with {AI} overviews.
\newblock \url{https://tinyurl.com/mr2zme39}, August 2024.
\newblock Accessed: 2025-12-19.

\bibitem[Burtch et~al.(2024)Burtch, Lee, and Chen]{burtch2024consequences}
G.~Burtch, D.~Lee, and Z.~Chen.
\newblock The consequences of generative ai for online knowledge communities.
\newblock \emph{Scientific Reports}, 14\penalty0 (1):\penalty0 10413, 2024.

\bibitem[Calzada and Gil(2020)]{calzada2020news}
J.~Calzada and R.~Gil.
\newblock What do news aggregators do? evidence from google news in spain and germany.
\newblock \emph{Marketing Science}, 39\penalty0 (1):\penalty0 134--167, 2020.

\bibitem[Cameron et~al.(2011)Cameron, Gelbach, and Miller]{Cameron2011}
A.~C. Cameron, J.~B. Gelbach, and D.~L. Miller.
\newblock Robust inference with multiway clustering.
\newblock \emph{Journal of Business \& Economic Statistics}, 29\penalty0 (2):\penalty0 238--249, 2011.

\bibitem[Champion and Hill(2025)]{champion2025countering}
K.~Champion and B.~M. Hill.
\newblock Countering underproduction of peer produced goods.
\newblock \emph{New Media \& Society}, 27\penalty0 (9):\penalty0 5132--5150, 2025.

\bibitem[Chapekis and Lieb(2025)]{Pew2025AISummariesClicks}
A.~Chapekis and A.~Lieb.
\newblock Google users are less likely to click on links when an {AI} summary appears in the results.
\newblock \emph{Pew Research Center}, July 2025.

\bibitem[Chiou and Tucker(2017)]{chiou2017content}
L.~Chiou and C.~Tucker.
\newblock Content aggregation by platforms: The case of the news media.
\newblock \emph{Journal of Economics \& Management Strategy}, 26\penalty0 (4):\penalty0 782--805, 2017.

\bibitem[Davies(2025)]{Digiday2025AIO}
J.~Davies.
\newblock Google ai overviews linked to 25\% drop in publisher referral traffic, new data shows.
\newblock \url{https://tinyurl.com/4t7bfs99}, August 2025.

\bibitem[Davies and Guaglione(2026)]{Digiday2026AIOOptOut}
J.~Davies and S.~Guaglione.
\newblock Google's {AI} opt-out leaves publishers with a choice they can't safely use.
\newblock \url{https://digiday.com/media/googles-ai-opt-out-leaves-publishers-with-a-choice-they-cant-safely-use/}, June 2026.

\bibitem[del Rio-Chanona et~al.(2024)del Rio-Chanona, Laurentsyeva, and Wachs]{del2024large}
R.~M. del Rio-Chanona, N.~Laurentsyeva, and J.~Wachs.
\newblock Large language models reduce public knowledge sharing on online q\&a platforms.
\newblock \emph{PNAS nexus}, 3\penalty0 (9):\penalty0 pgae400, 2024.

\bibitem[Edmonds(2025)]{edmonds2025reddit}
L.~Edmonds.
\newblock {Reddit Is the 2nd Most-Cited Source in Google AI Overviews, but That Might Not Mean Much for Its Bottom Line}.
\newblock \url{https://tinyurl.com/4b78f2mu}, June 2025.
\newblock Yahoo Finance; originally published by Business Insider. Accessed July 18, 2026.

\bibitem[Fletcher and Nielsen(2018)]{fletcher2018automated}
R.~Fletcher and R.~K. Nielsen.
\newblock Automated serendipity: The effect of using search engines on news repertoire balance and diversity.
\newblock \emph{Digital Journalism}, 6\penalty0 (8):\penalty0 976--989, 2018.

\bibitem[Goldfarb and Tucker(2019)]{goldfarb2019digital}
A.~Goldfarb and C.~Tucker.
\newblock Digital economics.
\newblock \emph{Journal of economic literature}, 57\penalty0 (1):\penalty0 3--43, 2019.

\bibitem[Google(2025)]{Google2025AdsInAIO}
Google.
\newblock More opportunities for your business on google search.
\newblock \url{https://tinyurl.com/2snbhp7x}, May 2025.
\newblock Accessed: 2026-01-15.

\bibitem[{Google Ad Manager Help}({n.d.})]{GoogleAdManagerFillRate}
{Google Ad Manager Help}.
\newblock Ad manager report metrics.
\newblock \url{https://tinyurl.com/5s38x8my}, {n.d.}
\newblock Defines total fill rate as total impressions divided by total ad requests. Accessed July 15, 2026.

\bibitem[Greenstein(2010)]{greenstein2010gatekeeping}
S.~Greenstein.
\newblock Gatekeeping economics.
\newblock \emph{IEEE Micro}, 30\penalty0 (5):\penalty0 102--104, 2010.

\bibitem[Harsel et~al.(2025)Harsel, Drozdov, and Skopec]{SemrushMostCitedDomainsAI2025}
L.~Harsel, A.~Drozdov, and C.~Skopec.
\newblock The most-cited domains in ai: A 3-month study.
\newblock \url{https://tinyurl.com/463t7evc}, 2025.
\newblock Semrush blog post, published November 10, 2025.

\bibitem[{Interactive Advertising Bureau (IAB)}(2025)]{IAB2025AdRevenue2024}
{Interactive Advertising Bureau (IAB)}.
\newblock Digital ad revenue surges 15\% yoy in 2024, climbing to \$259b, according to iab.
\newblock IAB news release (conducted by PwC), April 2025.
\newblock Accessed 2025-12-20.

\bibitem[Khromova and Ivanna(2024)]{seranking2024keywords}
Y.~Khromova and V.~Ivanna.
\newblock {AI} overviews research: Comparing pre and post-rollout results on {100}k keywords.
\newblock \url{https://tinyurl.com/3z59d95a}, 2024.

\bibitem[Lemmerich et~al.(2019)Lemmerich, S{\'a}ez-Trumper, West, and Zia]{lemmerich2019world}
F.~Lemmerich, D.~S{\'a}ez-Trumper, R.~West, and L.~Zia.
\newblock Why the world reads wikipedia: Beyond english speakers.
\newblock In \emph{Proceedings of the twelfth ACM international conference on web search and data mining}, pages 618--626, 2019.

\bibitem[Lyu et~al.(2025)Lyu, Siderius, Li, Acemoglu, Huttenlocher, and Ozdaglar]{lyu2025wikipedia}
L.~Lyu, J.~Siderius, H.~Li, D.~Acemoglu, D.~Huttenlocher, and A.~Ozdaglar.
\newblock Wikipedia contributions in the wake of chatgpt.
\newblock In \emph{Companion Proceedings of the ACM on Web Conference 2025}, pages 1176--1179, 2025.

\bibitem[McConnell(2024)]{mcconnell2024can}
B.~McConnell.
\newblock Can’t see the forest for the logs: on the perils of using difference-in-differences with a log-dependent variable, 2024.

\bibitem[Miller(2025)]{Wikimedia2025traffic}
M.~Miller.
\newblock New user trends on wikipedia.
\newblock \url{https://tinyurl.com/3r8k9ysa}, October 2025.

\bibitem[Miz et~al.(2020)Miz, Hanna, Aspert, Ricaud, and Vandergheynst]{miz2020trending}
V.~Miz, J.~Hanna, N.~Aspert, B.~Ricaud, and P.~Vandergheynst.
\newblock What is trending on wikipedia? capturing trends and language biases across wikipedia editions.
\newblock In \emph{Companion proceedings of the Web conference 2020}, pages 794--801, 2020.

\bibitem[Nielsen and Fletcher(2022)]{Nielsen2022Concentration}
R.~K. Nielsen and R.~Fletcher.
\newblock Concentration of online news traffic and publishers' reliance on platform referrals: Evidence from passive tracking data in the uk.
\newblock \emph{Journal of Quantitative Description: Digital Media}, 2, 2022.

\bibitem[Padilla et~al.(2025)Padilla, Lam, Lambrecht, and Hollenbeck]{padilla2025impact}
N.~Padilla, H.~T. Lam, A.~Lambrecht, and B.~Hollenbeck.
\newblock The impact of llm adoption on online user behavior.
\newblock \emph{Available at SSRN}, 2025.

\bibitem[Piccardi et~al.(2023)Piccardi, Gerlach, Arora, and West]{piccardi2023large}
T.~Piccardi, M.~Gerlach, A.~Arora, and R.~West.
\newblock A large-scale characterization of how readers browse wikipedia.
\newblock \emph{ACM Transactions on the Web}, 17\penalty0 (2):\penalty0 1--22, 2023.

\bibitem[{Playwire}(2026)]{Playwire2026StatePublisherAdRevenue}
{Playwire}.
\newblock State of publisher ad revenue 2026: Insights and opportunities.
\newblock Technical report, Playwire, 2026.
\newblock URL \url{https://tinyurl.com/2s32r2rd}.

\bibitem[Reid(2023)]{Google2024SuperChargeSearch}
E.~Reid.
\newblock Supercharging search with generative {AI}.
\newblock \url{https://tinyurl.com/yc5xkbm3}, May 2023.

\bibitem[Reid(2024{\natexlab{a}})]{Google2024AIOUpdate}
E.~Reid.
\newblock {AI} overviews: About last week.
\newblock \url{https://blog.google/products-and-platforms/products/search/ai-overviews-update-may-2024/}, May 2024{\natexlab{a}}.

\bibitem[Reid(2024{\natexlab{b}})]{Google2024GenerativeAISearch}
E.~Reid.
\newblock Generative {AI} in search: Let {Google} do the searching for you.
\newblock \url{https://tinyurl.com/4y3ehssc}, May 2024{\natexlab{b}}.

\bibitem[Reid(2025)]{Google2025AISearchClicks}
E.~Reid.
\newblock {AI} in search: Driving more queries and higher quality clicks.
\newblock \url{https://tinyurl.com/ykk2ucf2}, August 2025.

\bibitem[{Reuters}(2026)]{Reuters2026GoogleAISummariesLawsuit}
{Reuters}.
\newblock Google defends ai search summaries against publishers’ lawsuit.
\newblock \url{https://tinyurl.com/43zy2abk}, Jan. 2026.
\newblock Reuters Legal, accessed January 2026.

\bibitem[Schwartz(2025)]{SearchEngineLand2025EUAIOverviews}
B.~Schwartz.
\newblock {Google Rolls Out AI Overviews In EU Regions}.
\newblock \url{https://searchengineland.com/google-rolls-out-ai-overviews-in-eu-regions-453595}, Mar. 2025.
\newblock Search Engine Land. Accessed August 22, 2026.

\bibitem[SemRush(2025)]{SemRushWikipedia}
SemRush.
\newblock Wikipedia traffic stats.
\newblock \url{https://www.semrush.com/website/wikipedia.org/overview/}, 2025.

\bibitem[Similarweb(2025)]{SimilarwebWikipedia}
Similarweb.
\newblock Wikipedia website performance.
\newblock \url{https://www.similarweb.com/website/wikipedia.org/}, 2025.

\bibitem[{Similarweb}(2026)]{Similarweb}
{Similarweb}, 2026.

\bibitem[{Statista}(2025)]{StatistaSearchShare}
{Statista}.
\newblock Market share of leading search engines worldwide from january 2015 to december 2025.
\newblock \url{https://tinyurl.com/f46smwst}, 2025.
\newblock Accessed: 2026-01-19.

\bibitem[{Statistics}(2026)]{statistics2026google}
{Statistics}.
\newblock Distribution of search engine referrals share worldwide from january {2015} to june {2026}.
\newblock \url{https://www.statista.com/statistics/1381664/worldwide-all-devices-market-share-of-search-engines/?srsltid=AfmBOoqaLLZ-L0fBrBd0tOF4gVFl4rCKI1Ce4OfnQVGCavUDlS7IFLJR}, 2026.

\bibitem[Stier et~al.(2020)Stier, Kirkizh, Froio, and Schroeder]{stier2020populist}
S.~Stier, N.~Kirkizh, C.~Froio, and R.~Schroeder.
\newblock Populist attitudes and selective exposure to online news: A cross-country analysis combining web tracking and surveys.
\newblock \emph{The International Journal of Press/Politics}, 25\penalty0 (3):\penalty0 426--446, 2020.

\bibitem[Tabuena(2025)]{Tabuena2025OptimizeAIO}
E.~J. Tabuena.
\newblock New ahrefs data outline how to optimize for ai overviews.
\newblock \url{https://tinyurl.com/2kwfy95z}, 2025.
\newblock DesignRush News. Accessed: 2026-01-19.

\bibitem[Venkatachary(2024)]{Google2024AIOExpansion}
S.~Venkatachary.
\newblock {AI} overviews in search are coming to more places around the world.
\newblock \url{https://tinyurl.com/mrx9xfkc}, Octobor 2024.

\bibitem[{Wikimedia Foundation}(2026)]{WikimediaClickstream}
{Wikimedia Foundation}.
\newblock {Wikipedia Clickstream}.
\newblock \url{https://meta.wikimedia.org/wiki/Research:Wikipedia_clickstream}, 2026.
\newblock Documentation for the monthly public clickstream releases. Accessed July 31, 2026.

\bibitem[Wikipedia(2024)]{WikipediaTopLanguages}
Wikipedia.
\newblock Languages used on the internet.
\newblock \url{https://tinyurl.com/ynwy2ajc}, 2024.

\bibitem[Wikipedia(2025{\natexlab{a}})]{TopVisited2024}
Wikipedia.
\newblock List of most-visited websites.
\newblock \url{https://tinyurl.com/4css9t5x}, 2025{\natexlab{a}}.
\newblock Accessed: 22 January 2026.

\bibitem[Wikipedia(2025{\natexlab{b}})]{WikipediaStats2024}
Wikipedia.
\newblock Wikipedia:statistics.
\newblock \url{https://tinyurl.com/y32j75p6}, 2025{\natexlab{b}}.
\newblock Accessed: 22 January 2026.

\bibitem[Wooldridge(2010)]{wooldridge2010econometric}
J.~M. Wooldridge.
\newblock \emph{Econometric analysis of cross section and panel data}.
\newblock MIT press, 2010.

\bibitem[Yan et~al.(2025)Yan, Zhong, Zhong, Zhou, and Mehta]{yan2025impact}
S.~Yan, H.~Zhong, Z.~Z. Zhong, W.~Zhou, and N.~Mehta.
\newblock Reasoning {AI}, consumer search, and purchase: Evidence from an online platform field experiment.
\newblock Working Paper 5631050, SSRN, 2025.
\newblock URL \url{https://ssrn.com/abstract=5631050}.

\bibitem[Ye and Yoganarasimhan(2025)]{YeYoganarasimhan2025}
Z.~Ye and H.~Yoganarasimhan.
\newblock Fair document valuation in llm summaries via shapley values.
\newblock \emph{arXiv preprint arXiv:2505.23842}, 2025.

\bibitem[Zeng et~al.(2020)Zeng, Kohno, and Roesner]{ZengKohnoRoesner2020BadNews}
E.~Zeng, T.~Kohno, and F.~Roesner.
\newblock Bad news: Clickbait and deceptive ads on news and misinformation websites.
\newblock In \emph{Workshop on Technology and Consumer Protection (ConPro '20)}, 2020.
\newblock Accessed: 2026-01-19.

\bibitem[Zhang et~al.(2026)Zhang, Cui, and Zhang]{zhang2026impact}
P.~Zhang, R.~Cui, and D.~J. Zhang.
\newblock The impact of ai search on the online content ecosystem: Evidence from google and reddit.
\newblock \emph{arXiv preprint arXiv:2605.16428}, 2026.

\bibitem[Zhang and Zhu(2011)]{zhang2011group}
X.~Zhang and F.~Zhu.
\newblock Group size and incentives to contribute: A natural experiment at chinese wikipedia.
\newblock \emph{American Economic Review}, 101\penalty0 (4):\penalty0 1601--1615, 2011.

\bibitem[Zheng et~al.(2025)Zheng, Zhu, Ye, and Shen]{ZhengEtAl2025GenerativeSearch}
S.~Zheng, Y.~Zhu, X.~Ye, and L.~Shen.
\newblock Generative search: Evidence from a large-scale field experiment.
\newblock Working Paper 5297194, SSRN, 2025.
\newblock URL \url{https://ssrn.com/abstract=5297194}.

\end{thebibliography}


\begin{thebibliography}{6}
\providecommand{\natexlab}[1]{#1}
\providecommand{\url}[1]{\texttt{#1}}
\expandafter\ifx\csname urlstyle\endcsname\relax
  \providecommand{\doi}[1]{doi: #1}\else
  \providecommand{\doi}{doi: \begingroup \urlstyle{rm}\Url}\fi

\bibitem[{Association of National Advertisers (ANA)}(2025)]{ANA2025ProgrammaticBenchmarkQ1}
{Association of National Advertisers (ANA)}.
\newblock Programmatic transparency benchmark report (q1 2025).
\newblock \url{https://tinyurl.com/4jc4r2pm}, 2025.

\bibitem[Budaraju(2024)]{Google2024AIOAugExpansion}
H.~Budaraju.
\newblock New ways to connect to the web with {AI} overviews.
\newblock \url{https://tinyurl.com/mr2zme39}, August 2024.
\newblock Accessed: 2025-12-19.

\bibitem[{Google Ad Manager Help}({n.d.})]{GoogleAdManagerFillRate}
{Google Ad Manager Help}.
\newblock Ad manager report metrics.
\newblock \url{https://tinyurl.com/5s38x8my}, {n.d.}
\newblock Defines total fill rate as total impressions divided by total ad requests. Accessed July 15, 2026.

\bibitem[{Playwire}(2026)]{Playwire2026StatePublisherAdRevenue}
{Playwire}.
\newblock State of publisher ad revenue 2026: Insights and opportunities.
\newblock Technical report, Playwire, 2026.
\newblock URL \url{https://tinyurl.com/2s32r2rd}.

\bibitem[Rambachan and Roth(2023)]{rambachan2023more}
A.~Rambachan and J.~Roth.
\newblock A more credible approach to parallel trends.
\newblock \emph{Review of Economic Studies}, 90\penalty0 (5):\penalty0 2555--2591, 2023.

\bibitem[Zeng et~al.(2020)Zeng, Kohno, and Roesner]{ZengKohnoRoesner2020BadNews}
E.~Zeng, T.~Kohno, and F.~Roesner.
\newblock Bad news: Clickbait and deceptive ads on news and misinformation websites.
\newblock In \emph{Workshop on Technology and Consumer Protection (ConPro '20)}, 2020.
\newblock Accessed: 2026-01-19.

\end{thebibliography}
\end{document}